\documentclass[letterpaper,10pt,aps,prx,twocolumn,floatfix,superscriptaddress]{revtex4-2}
\usepackage{amsmath}
\usepackage{amssymb}
\usepackage{mathtools}
\usepackage{graphicx}
\usepackage{booktabs}
\usepackage{longtable}
\usepackage{multirow}
\usepackage{float}
\usepackage{color}
\usepackage{comment}
\usepackage[english]{babel}
\babelprovide[hyphenrules=english]{en}

\newcommand{\ket}[1]{|#1\rangle}
\newcommand{\bra}[1]{\langle#1|}
\newcommand{\ketbra}[2]{\ket{#1}\!\bra{#2}}
\newcommand{\ii}{\mathrm{i}}
\newcommand{\kk}{\mathbf k}
\newcommand{\LL}{\mathbf L}
\newcommand{\QQ}{\mathbf Q}
\newcommand{\RR}{\mathbf R}
\newcommand{\rr}{\mathbf r}
\newcommand{\NTHC}{N_{\mathrm{THC}}}
\newcommand{\NAO}{N_{\mathrm{AO}}}

\begin{document}

\title{Quasiparticle quantum simulation of materials with the Bethe-Salpeter equation}

\begin{abstract}
The first-principles quantum simulation of the full electronic Hamiltonian of a material model of realistic size appears daunting even in a future era of fully fault-tolerant quantum computers.
In contrast, classical simulations of material excited states rely on heuristic quasiparticle frameworks, such as the GW approximation and Bethe-Salpeter equation, where only a few quasiparticles are explicitly considered.
Here we give a detailed description of quasiparticle quantum simulation for multi-excitonic states within the BSE.
We develop an explicit first-quantized, orbital-based block encoding of the multi-exciton BSE Hamiltonian for qubitization, and reduce its cost using integral factorization and crystal symmetries.
For a materials model of size $N$, and a fixed number of quasiparticles, the resulting algorithms require $\widetilde{O}(N)$ Toffoli gates and $O(\log N)$ logical qubits, or, using space-time tradeoffs, $\widetilde{O}(\sqrt{N})$ Toffoli gates and $\widetilde{O}(\sqrt{N})$ logical qubits, achieving up to a sixth power reduction in Toffoli count, or up to exponential reduction in qubit count, versus standard full Hamiltonian simulation techniques that have been explored in materials simulation.
Resource estimates for the classically challenging problem of triexcitonic eigenvalue estimation across a range of semiconductor models of a realistic physical size of $O(10^3)$ atoms, show that our Toffoli-optimized implementation reduces the Toffoli count by $10^6-10^7$ relative to full Hamiltonian simulation estimates, while a qubit-optimized implementation requires only a few hundred logical qubits.
More broadly, these results show that classical heuristic frameworks will be essential in the design of future fault-tolerant quantum algorithms for materials.
\end{abstract}

\author{Jielun Chen}
\thanks{These authors contributed equally to this work.}
\affiliation{Division of Physics, Mathematics, and Astronomy, California Institute of Technology, Pasadena, California 91125, USA}
\affiliation{Institute for Quantum Information and Matter, Pasadena, California 91125, USA}
\affiliation{Marcus Center for Theoretical Chemistry, California Institute of Technology, Pasadena, California 91125, USA}
\author{Jiace Sun}
\thanks{These authors contributed equally to this work.}
\affiliation{Division of Chemistry and Chemical Engineering, California Institute of Technology, Pasadena, California 91125, USA}
\affiliation{Marcus Center for Theoretical Chemistry, California Institute of Technology, Pasadena, California 91125, USA}
\author{Garnet Kin-Lic Chan}
\email[Corresponding author: ]{gkc1000@gmail.com}
\affiliation{Division of Chemistry and Chemical Engineering, California Institute of Technology, Pasadena, California 91125, USA}
\affiliation{Institute for Quantum Information and Matter, Pasadena, California 91125, USA}
\affiliation{Marcus Center for Theoretical Chemistry, California Institute of Technology, Pasadena, California 91125, USA}

\maketitle

\section{Introduction}

The quantum simulation of materials is a highly anticipated application of quantum computing~\cite{bauer2020quantum,mcardle2020quantum,bauer2016hybrid,babbush2018lowdepth,kivlichan2020improved,ma2020quantum,delgado2022simulating,zini2023quantum,rubin2023faulttolerant,berry2024quantum,rubin2024quantum,clinton2024towards}.
For many materials, the key questions involve the electronic structure, spectroscopy, and dynamics.
These can be addressed by solving the time-independent and time-dependent electronic Schr\"odinger equations.

In the context of first principles fault-tolerant quantum materials simulation, computing the electronic Hamiltonian eigenstates and dynamics has already been explored~\cite{babbush2018lowdepth,su2021faulttolerant,ivanov2023quantum,rubin2023faulttolerant,berry2024quantum}.
These formulations have focused on exact simulations of the full electronic Hamiltonian of the material.
However, since current first principles material simulations with density functional theory often involve a few thousand atoms and $O(10^4)$ electrons or more, replacing such classical simulations is a formidable task.
As we detail below, using existing techniques of full Hamiltonian simulation, the Toffoli counts needed to compute an excitation energy to a relevant precision exceed even the resources that are expected to be available in the era of full fault-tolerance by many orders of magnitude.
Additional approximations must therefore be made for the quantum simulation of materials to be practical.

Full Hamiltonian simulation is not the problem that is tackled by most beyond density functional classical approaches.
Instead, these work within a quasiparticle picture of electronic behaviour.
This limits the explicit modeling of correlation effects to a few (quasi)particles.
While the quasiparticle equations are augmented with additional heuristic assumptions that mean that they are not exact, their domain of applicability is understood through empirical validation.
For example, the GW and GW/Bethe-Salpeter quasiparticle theories are probably the most commonly applied electronic structure simulations in materials beyond density functional theory~\cite{hedin1965new,hybertsen1986electron,onida2002electronic,golze2019gw,blase2018bethe}, and have found particular value in the modeling of semiconductor excitations~\cite{rohlfing2000electronhole}.

Recently, we have described the formulation of quantum simulation within the quasiparticle approximation and have shown that, in a setting that takes advantage of low-energy quasiparticles as well as locality, and under standard initial state preparation assumptions, quantum algorithms can achieve a large asymptotic speedup over classical algorithms within the same heuristic framework~\cite{chen2025framework}.
Here, we move beyond asymptotic scaling and provide a detailed analysis of the quasiparticle GW/BSE for excited state properties and dynamics.
We focus only on the quasiparticle heuristic rather than also using locality, as this more general setting covers a wider range of materials, and provides a baseline for the incorporation of more heuristic information in the future.
We focus on the modeling of high order multi-excitonic states and dynamics in semiconductors, which is extremely expensive and which has yet to be studied in first principles classical simulations.
We achieve, relative to full Hamiltonian quantum simulation approaches that have been benchmarked in materials, a $\sim 6$th order polynomial speedup in scaling when optimizing for Toffoli count, and up to an exponential compression when optimizing for qubit count.
Relative to standard classical matrix-vector algorithms for the Bethe-Salpeter equation, the speedup is exponential in the multi-exciton order $m$, and for the tri-excitonic case, corresponds to a $\sim 12$th order speedup in speed and an exponential reduction in classical resources.

We perform explicit compilation of quasiparticle quantum simulation, working within a first-quantized, orbital basis formulation.
Our compilation has some features in common with that of first- and second-quantized Hamiltonian simulation compilation, such as the use of factorized integral representations, qubit/Toffoli data loading tradeoffs, and symmetry adaptation, but we use a new block encoding strategy that is adapted to the physical form of the GW/BSE Hamiltonian.
We carry out quantum resource estimates of tri-excitonic state simulation for a range of semiconductor materials, where we find {$\sim 10^6-10^7$} reduction in Toffoli-gate count relative to full Hamiltonian simulation using realistic basis sets and problem sizes of a few thousand atoms.
Our work therefore provides quantitative evidence of the advantages of quasiparticle quantum simulation in future first-principles studies of materials using quantum computers.

\section{Quasi-particle quantum simulation}

\subsection{The Bethe-Salpeter equation}

We are concerned with simulating electronic structure and dynamics.
In atomic units, the corresponding electronic Hamiltonian in the non-relativistic approximation is
\begin{align}
  H = \sum_i -\frac{1}{2}\nabla_i^2 + v_{\text{ext}}(\mathbf{r}_i) + \sum_{i>j} \frac{1}{r_{ij}}\label{eq:Hamiltonians}
\end{align}
where for an isolated system $v_\text{ext}(\mathbf{r})$ is the external potential from the nuclei.
The system size will be denoted by $N$, and we generally consider the setting where the number of electrons is $\propto N$.
The task of quantum simulation is to obtain the eigenstates and dynamics given by the time-independent and dependent Schr\"odinger equations
\begin{align}
  {H} \Psi = E \Psi \\
  i \partial_t \Psi = {H} \Psi
\end{align}

One common framework is to simulate in second-quantization, where a single particle orthogonal basis $\{ \phi_i\}$ is introduced, and ${H}$ is approximated by its finite basis second quantized counterpart~\cite{mcardle2020quantum,bauer2020quantum}.
Another framework is first quantized quantum simulation~\cite{su2021faulttolerant}.
There, the computational basis states are simple product states, and antisymmetrization is enforced by initial state preparation.
Both first- and second-quantized approaches have been studied in materials simulation, and their relative advantages and disadvantages discussed~\cite{berry2024quantum,rubin2023faulttolerant}.

The above frameworks simulate all electrons in a material.
However, in a macroscopic sample, this may approach $\sim 10^{23}$ electrons.
While most electronic properties converge at smaller system sizes, as a practical guide we can consider the problem sizes in density functional materials calculations, which in simple semiconductors typically involve $O(1000)$ atoms, with $\sim 10^4$ electrons or more.
These would require impractical amounts of quantum resources to address.

However, in many materials of interest, the low-energy excitations are captured by a small number ($m_{\mathrm{tot}}=O(1)$ in total) of electronic quasiparticles, namely, dressed electrons and holes.
This suggests that one may not need to simulate the vast majority of the electrons.
Formally, quasiparticles are described by Green's functions~\cite{martin2016interacting}.
For example, writing $1\equiv(\mathbf{r}_{1}\sigma_{1}t_{1})$ (where $\sigma$ is the spin index) and $\hat\psi(1)$ for the Heisenberg-picture field operator, then individual quasiparticles, and pairs of quasiparticles, are described by the zero-temperature 1- and 2-particle Green's functions
\begin{align}
  G(1,1') &= -i\,\bigl\langle\Psi_{0}\bigr|\,
  \mathcal{T}\bigl[\hat\psi(1)\,\hat\psi^{\dagger}(1')\bigr]\,
  \bigl|\Psi_{0}\bigr\rangle,
  \\
  G_{2}(1,2;1',2') &= (-i)^{2}\bigl\langle\Psi_{0}\bigr|\,
  \mathcal{T}\bigl[\hat\psi(1)\hat\psi(2)\,
  \hat\psi^{\dagger}(2')\hat\psi^{\dagger}(1')\bigr]\,
  \bigl|\Psi_{0}\bigr\rangle,
\end{align}
where $\mathcal{T}$ is the time-ordering operator and $|\Psi_{0}\rangle$ is the ground state of ${H}$; we abbreviate $G\equiv G_{1}$.
We can write down equations-of-motion for the Green's functions, such as the Dyson equation for $G$,
\begin{equation}
  i\partial_{t_{1}}G(1,1')=\delta(1,1')+h(1)G(1,1')
  +\int\! d2\;\Sigma(1,2)\,G(2,1'),
  \label{eq:dyson}
\end{equation}
where $\Sigma$ is the self-energy that encapsulates all the many-body effects.

Limiting ourselves to neutral excitations for simplicity, we denote the number of electron--hole pairs by $m$, so that $m_{\mathrm{tot}}=2m$.
An excitation with $m$ electrons and $m$ holes is described by $G_{2m}$; for $m=1$, the excitation is a `single' excitonic state (note, we use excitonic state to refer to any excitation, regardless of whether the electron and hole are bound; the bound-case corresponds to an exciton) and for $m>1$, it is a multi-excitonic state.
$G_{2m}$ captures the exact (multi)-excitonic energies and dynamics.
However, solving the exact equation of motion is no easier than solving for the full dynamics, due to the complexity of the exact self-energy and similar quantities.
Thus, in practical quasiparticle classical simulations, additional approximations are made.
The validity and accuracy of these heuristic approximations (for certain applications) is ultimately justified by their empirical success and widespread use.
From a quantum simulation perspective, these approximations are also significant: the exact equations of motion are non-linear and involve non-Hermitian operators, while the standard classical approximations render them linear with Hermitian operators.

Although we are ultimately interested in the quantum simulation of multi-excitonic states and dynamics, we first present the simplification of the equation-of-motion for single-excitonic states to reduce the notational complexity.
Single excitonic energies appear as poles of the particle-hole correlation function, written as
\begin{equation}
  L(1,2;1',2')=-G_{2}(1,2;1',2')+G(1,1')G(2,2'),
  \label{eq:L_definition}
\end{equation}
The equation-of-motion of $L$ is the Bethe-Salpeter equation (BSE)~\cite{salpeter1951relativistic,strinati1988application},
\begin{widetext}
\begin{align}
  L(1,2;1',2')=L_{0}(1,2;1',2')
  +\int\! d(3456)\;L_{0}(1,4;1',3)\,K(3,5;4,6)\,L(6,2;5,2'),
  \label{eq:bse}
\end{align}
\end{widetext}
where $L_{0}=G(1,2')G(2,1')$.

We next introduce the approximate GW-BSE expressions that are used when solving the BSE in first-principles calculations~\cite{rohlfing2000electronhole,onida2002electronic}.
Within this formulation, $\Sigma \approx G \cdot W$~\cite{hedin1965new,hybertsen1986electron} where $\cdot$ abbreviates the integrations, $W$ is the screened interaction, and further, in the subsequent BSE equation, $\Sigma$ is taken to be static.
Then single quasiparticles are eigenstates of a modified single-particle Hamiltonian, written as $\sum_i \epsilon_p |p\rangle \langle p|$, where $p$ labels the single particle state, and $\epsilon_i$ is its (self-energy) corrected eigenvalue.
Next, the kernel $K(3,5;4,6)$ is approximated by
\begin{align}
  K(3,5;4,6) = -i \delta(3,4) \delta(5,6) V(3,6) + i \delta(3,6) \delta(4,5) W(3,4)
\end{align}
The first term involves the standard Coulomb interaction $V(3,6)=1/|\mathbf{r}_3-\mathbf{r}_6|$, this term is called the exchange term.
The second term involves the screened Coulomb interaction $W(3,4)$ and is called the direct term; this is formally frequency dependent also, but in most GW-BSE calculations, one uses the static approximation $W(3,4) = W(\mathbf{r}_3,\mathbf{r}_4,\omega=0)$.
Finally, dropping the coupling between forwards-in-time and backwards-in-time propagating electron-hole pairs (the Tamm-Dancoff approximation), we arrive at the standard GW-BSE description of dynamics where the equation-of-motion of $L$ is generated by a static Hermitian electron-hole effective Hamiltonian ${H}_\mathrm{BSE}$,
\begin{align}
  {H}_\text{BSE} = {H}_0 + {K}
\end{align}
We can write the two parts of ${K}$ as $H_{ov}^\text{ex} + H_{ov}^\text{dir}$; the exchange part describes the scattering from one electron-hole pair into another one by the Coulomb interaction, while the direct part describes the scattering between the electron and hole in an electron-hole pair via the screened interaction.
The determination of eigenstates and dynamics within the Bethe--Salpeter equation reduces to the solution of a time-independent and time-dependent Schr\"odinger equation,
\begin{align}
  H_{\text{BSE}}A&=\Omega A,\label{eq:bse_eigen}\\
  i\partial_{t}A&=H_{\text{BSE}}A.\label{eq:bse_dynamics}
\end{align}
where $A$ is the excitonic state wavefunction and $\Omega$ is the excitation energy.

We now see that we can make a simple generalization of GW-BSE to multi-excitonic states.
This is achieved by taking the Hilbert space to be that of multi-excitonic states, and describing the propagation of electrons and holes by the Fock term $H_0$ and the scattering between electron-hole pairs by $H_{ov}^\text{ex}$ and $H_{ov}^\text{dir}$.
The only additional ingredient is to treat the electron-electron and hole-hole interactions between different excitons; this takes place via the screened interaction, similar to the direct term.

\subsection{General nature of speedups versus classical and standard quantum simulation}

We may now choose to solve the GW-BSE quasiparticle Schrodinger equations with standard quantum algorithms, for example quantum phase estimation for the eigenvalue problem and Hamiltonian simulation for the dynamics problems.
The main difference from the standard quantum simulation problem is that it now involves only $m_{\mathrm{tot}}$ quasiparticles, counting electrons and holes together.
The Hilbert space is polynomial in size and thus the classical GW-BSE simulation methods are polynomial cost in $N$, which is why they are widely used.
The quantum speedup we can expect is therefore a polynomial one.
We can now establish the origin of the quantum versus classical speedup and the speedup of quasiparticle quantum algorithms versus standard full Hamiltonian simulation quantum algorithms.

The classical linear algebra algorithm involves linear algebra in a Hilbert space of dimension $N^{m_{\mathrm{tot}}}$.
Without additional heuristic structure, we store the classical vector with space $N^{m_{\mathrm{tot}}}$, and the cost is that of operating on it with the Hamiltonian operator.
{For a typical 2-body Hamiltonian whose integral matrix elements can be written as a $N^2 \times N^2$ matrix, the cost of applying the Hamiltonian by a matrix-vector product has a $O(N^{m_{\mathrm{tot}}+2})$ cost.}

In the quasiparticle quantum simulation, one is carrying out quantum linear algebra.
The qubit count exponentially compresses the Hilbert space, thus there is an exponential savings to represent the state.
In terms of gate counts, the cost is to implement the Hamiltonian as a block-encoding~\cite{low2019qubitization,gilyen2019quantum}, where the cost is related to the classical data that needs to be loaded and the resulting subnormalization of the Hamiltonian, i.e.
\begin{equation}
  C \sim \text{subnormalization cost} \times \text{data}
\end{equation} However, because the Hamiltonian only contains some fixed-body level of interactions (e.g. two-particle interactions) the amount of data that needs to be loaded is $\mathrm{poly}(N)$, independent of $m_{\mathrm{tot}}$.
Similarly, because the Hamiltonian is an effective Hamiltonian for $m_{\mathrm{tot}}$-particles (regardless of system size), then its subnormalization (lower bounded by the largest singular value of the Hamiltonian) is expected to be $\mathrm{poly}(m_{\mathrm{tot}})$, thus we get the reduction from a $N^{m_{\mathrm{tot}}+c}$ classical cost to $\mathrm{poly}(m_{\mathrm{tot}}) \mathrm{poly}(N)$ quantum cost.

Similarly, we can compare an equivalent operation with the full quantum simulation (either in first or second quantization)~\cite{berry2024quantum,rubin2023faulttolerant}.
In this case, the cost has a similar form, the amount of data to load to represent the two-particle interactions is similarly $\mathrm{poly}(N)$.
However, since the Hamiltonian is now describing $N$-particles, we expect the subnormalization to be $\mathrm{poly}(N)$, thus giving a $\mathrm{poly}(N)/\mathrm{poly}(m_{\mathrm{tot}})$ factor speedup.

While we can tune these speedups to be arbitrarily large by making $m_{\mathrm{tot}}$ or $N$ arbitrarily large, in this work we will be interested in speedup under practical simulation settings.
This means that both $m_{\mathrm{tot}}$ and $N$ must correspond to physically relevant problem sizes, while the absolute costs must also be in a reasonable range for an actual simulation.
For material sizes we therefore consider $N$ in the range of $\sim 1000$ atoms.
For the neutral BSE applications considered here, biexcitons have $m=2$ electron--hole pairs ($m_{\mathrm{tot}}=4$), and triexcitons have $m=3$ pairs ($m_{\mathrm{tot}}=6$).
We mainly focus on the $m=3$ triexcitonic case in our resource estimates below as such states are generated and studied in physical experiments~\cite{bonati2005spectral,shulenberger2018multiexciton}, and serve as a balance between classical complexity and direct relevance in multi-excitonic physics.

\section{Explicit formulation}

\subsection{Computational representation}

We now specify the multi-exciton Bethe-Salpeter equations, within the Tamm-Dancoff formulation, and using static self-energies and static screening, as commonly used in the GW-BSE framework.
We assume we have obtained single-particle mean-field orbitals $\{ \phi \}$ and the self-energy corrected orbital energies $\{ \epsilon \}$.
We label the occupied (hole) orbitals by $i, j, \ldots$ and virtual (particle) orbitals, labelled by $a, b, \ldots$; when we do not distinguish between occupied and virtual indices, we use the labels $p, q, \ldots$.
The number of occupied labels will be denoted $N_o$, the number of virtual labels $N_v$, and the total number of labels $\NAO$.
The material contains $O(N)$ electrons, while the neutral $m$-exciton sector contains $m_{\mathrm{tot}}=2m$ quasiparticles ($m$ electrons and $m$ holes), and (without translational symmetry, see below) $\NAO = O(N)$, $N_o = O(N)$ and $N_v = O(N)$.
We will assume a spin-orbital basis, i.e. the basis states are of the form $|\phi \sigma\rangle$ with $\sigma \in \{ \uparrow, \downarrow \}$.
In a crystalline material, the occupied orbitals constitute the core and valence bands, while the virtual orbitals are the conduction bands.

The computational basis is the first-quantized mean-field $m$-particle/$m$-hole basis, i.e. the basis states are labelled
\begin{align}
  |ij\ldots ab \ldots) \equiv \phi_i(1) \phi_j(2) \ldots  \ \phi_a(1') \phi_b(2')
\end{align}
However, we only interested in the Hamiltonian in the antisymmetric sector of this space, i.e. spanned by states
\begin{align}
  |ij \ldots ab \ldots\rangle \equiv [\mathcal{A} \ \phi_i(1) \phi_j(2) \ldots] [\mathcal{A} \ \phi_a(1') \phi_b(2') \ldots]
\end{align}
where the two antisymmetrizers act separately on the hole and particle states, and the numeric label range is $1, 2, \ldots, m$.
Since we will always prepare antisymmetric initial states, we define $H_\mathrm{BSE}$ to obtain the correct matrix elements under this assumption, which leads to simplifications below (i.e. it removes the need for explicit exchange terms in $H_{oo}^{\mathrm{dir}}$ and $H_{vv}^{\mathrm{dir}}$).

The number of qubits to store the state is $O(m_{\mathrm{tot}}\log N)=O(m\log N)$, which is to be compared with classical dense vector storage of $O(N^{2m})$, and first- and second-quantized quantum simulation storage of $O(N\log N)$ and $O(N)$, respectively, when the electron number is proportional to $N$.
Thus the quasiparticle quantum representation uses exponentially less storage than the classical dense representation, as well as standard first- and second-quantized quantum simulation.

The multi-exciton BSE Hamiltonian is a sum of five terms
\begin{align}
  H_{\mathrm{BSE}} = H_0 + H_{vv}^{\mathrm{dir}} + H_{oo}^{\mathrm{dir}}
  + H_{ov}^{\mathrm{ex}} + H_{ov}^{\mathrm{dir}} .
  \label{eq:H_BSE}
\end{align}
(Similar forms of $H_{\mathrm{BSE}}$ appear in Refs.~\cite{steinhoff2018biexciton, torche2021biexcitons, torche2019firstprinciples, deilmann2016threeparticle}).
The Fock term $H_0$ propagates the electrons and holes.
We denote the occupied and virtual blocks of the effective single-particle Fock matrix by $f^o$ and $f^v$, respectively; here these blocks include the static self-energy correction.
In a general basis within each sector,
\begin{align}
  H_0 &= \sum_e\sum_{ab\sigma} f^v_{ab}\,
  \ketbra{a\sigma}{b\sigma}_e \nonumber\\
  &\quad-\sum_h\sum_{ij\sigma} f^o_{ji}\,
  \ketbra{i\sigma}{j\sigma}_h.
  \label{eq:H_0}
\end{align}
Here $h,e$ label the $m$ holes and electrons.
More commonly, one works in the orbital-energy basis, $f^o_{ij}=\epsilon_i\delta_{ij}$ and $f^v_{ab}=\epsilon_a\delta_{ab}$, so $H_0$ is diagonal.
The hole contribution carries $-(f^o)^{\mathsf T}$.
Thus $H_0$ denotes the full Fock term, whereas $f^o,f^v$ denote its single-particle blocks.

$H_{vv}^{\mathrm{dir}}$ and $H_{oo}^{\mathrm{dir}}$ describe the screened electron-electron and hole-hole interactions
\begin{align}
  H_{vv}^{\mathrm{dir}} &= \sum_{e<e'}\sum_{abcd}\sum_{\sigma,\sigma'} W_{ac,bd}\;
  \ketbra{a\,\sigma}{c\,\sigma}_{e}\,\ketbra{b\,\sigma'}{d\,\sigma'}_{e'},
  \label{eq:H_vv}\\
  H_{oo}^{\mathrm{dir}} &= \sum_{h<h'}\sum_{ijkl}\sum_{\sigma,\sigma'} W_{ki,l j}\;
  \ketbra{i\,\sigma}{k\,\sigma}_{h}\,\ketbra{j\,\sigma'}{l\,\sigma'}_{h_{'}} ,
  \label{eq:H_oo}
\end{align}

The direct term of $H_{ov}$
contains the screened electron-hole interaction \begin{align}
  H_{ov}^{\mathrm{dir}} = -\sum_{eh}\sum_{abij}\sum_{\sigma,\sigma'} W_{ba,ij}\;
  \ketbra{b\,\sigma}{a\,\sigma}_{e}\,\ketbra{j\,\sigma'}{i\,\sigma'}_{h},
  \label{eq:H_ov_dir}
\end{align}
where we note the minus sign, which produces exciton binding.

The remaining exchange term uses the bare Coulomb interaction in the $(ov|vo)$ ordering and exchanges the two spin labels
\begin{align}
  H_{ov}^{\mathrm{ex}} = \sum_{he}\sum_{abij}\sum_{\sigma,\sigma'} V_{ia,bj}\;
  \ketbra{j\,\sigma'}{i\,\sigma}_{h}\,\ketbra{b\,\sigma'}{a\,\sigma}_{e}.
  \label{eq:H_ov_ex}
\end{align}
This term is responsible for the energy splitting of singlet and triplet excitons.

The one and two-particle terms in $H_\mathrm{BSE}$ sum over $O(m)$ and $O(m^2)$ quasiparticle terms in their expressions, to be contrasted with the standard first quantized Hamiltonian, which sums over $O(N)$ and $O(N^2)$ terms.
This reflects the fact that we are working in a fixed $O(m)$ quasiparticle manifold.

\subsection{Integral representation and symmetry}
\label{sec:integrals}

The data specifying the Hamiltonian in Eq.~\eqref{eq:H_BSE} are the Fock blocks $f^o,f^v$ (or their eigenvalues $\epsilon$ in the orbital-energy basis) and the interaction integrals.
We use $X\in\{W,V\}$ for either the screened interaction $W$ or the bare Coulomb interaction $V$.
$W$ and $V$ have $O(N^4)$ entries and thus dominate the data requirements.
However, $W$ and $V$ are matrix elements of real-space interaction operators, and their data requirements can be reduced using tensor factorization~\cite{lee2021even,rubin2023faulttolerant}.
Here, we use tensor hypercontraction (THC)~\cite{hohenstein2012tensor,parrish2012tensor}, similarly to Ref.~\cite{lee2021even}, where we write
\begin{align}
  X_{pq,rs}\ &\approx\ \sum_{\mu\nu}
  \chi^{X*}_{p\mu}\,\chi^{X}_{q\mu}\;\zeta^{X}_{\mu\nu}\;
  \chi^{X*}_{r\nu}\,\chi^{X}_{s\nu},
  \label{eq:thc_W0}
\end{align}
where $\mu,\nu$ are auxiliary indices of dimension $\NTHC$.
The orbital indices $p,q,r,s$ range over both occupied and virtual orbitals; $i,j,\ldots$ and $a,b,\ldots$ select the occupied and virtual rows of the same tensor $\chi^X$.
We retain sector labels only for whole block operators, denoting these row restrictions by $\chi^{X,o}$ and $\chi^{X,v}$.
The $W$ and $V$ factorizations are fitted separately and may have different auxiliary dimensions; $\NTHC$ denotes the dimension of the factorization under discussion.
In standard electronic structure applications with a local atomic basis of dimension $\NAO$, the factorization is chosen to preserve a fixed relative error in the total energies by using constant $c_{\mathrm{IP}}=\NTHC/\NAO$ with system size.
This reduces the data to $O(N^2)$.

As analyzed in Ref.~\cite{rubin2023faulttolerant}, we can further reduce the data requirements using translational symmetry.
In a crystalline material, we can construct single-particle Bloch states labelled by crystal momentum $\mathbf{k}$.
We use the standard prescription of taking values on an evenly spaced Brillouin zone mesh with $N_k$ points~\cite{monkhorst1976special}; $N_k$ labels the size of an equivalent computational supercell with periodic boundary conditions, and for a fixed number of bands $\NAO$, $N = N_k \NAO$.
The crystalline electron and hole orbitals acquire are labelled by $i \mathbf{k}$ and $a \mathbf{k}$ respectively, where $N_o$ now labels the occupied bands per cell, and $N_v$, the virtual bands per cell.
Crystal-momentum conservation allows us to write the integrals as
\begin{equation}
  X_{p\kk,\,q(\kk\ominus\QQ),\,r(\kk'\ominus\QQ),\,s\kk'},
\end{equation}
where $\ominus$ denotes subtraction modulo reciprocal lattice vectors.
The common THC factorization is
\begin{align}
  &X_{p\kk,\,q(\kk\ominus\QQ),\,r(\kk'\ominus\QQ),\,s\kk'} \nonumber\\
  &\approx\frac{1}{N_k}\sum_{\mu\nu}
  \chi^{X*}_{p\kk,\mu}\,\chi^{X}_{q(\kk\ominus\QQ),\mu}\;
  \zeta^{X}_{\QQ,\mu\nu}\;
  \chi^{X*}_{r(\kk'\ominus\QQ),\nu}\,\chi^{X}_{s\kk',\nu}.
  \label{eq:thc_W}
\end{align}
This expression applies to all orbital sectors.
The direct terms use $X=W$ with the orbital-index choices in Eqs.~\eqref{eq:H_vv}--\eqref{eq:H_ov_dir}; the exchange term uses $X=V$ with the mixed occupied--virtual pairs in Eq.~\eqref{eq:H_ov_ex}.
The data size is then $O(N_k c_\mathrm{IP} \NAO^2) + O(N_k c_\mathrm{IP}^2 \NAO^2)$.
Since $\NAO$ and $c_\mathrm{IP}$ are independent of system size, and $N_k = O(N)$, the total data requirement is reduced to $O(N)$.
Beyond lattice translational symmetry, most crystals have additional symmetry elements given by their space group~\cite{bradley2010mathematical}.
We denote the finite symmetry group of the periodic supercell by $G$ and its compatible point group by $P$, so that $|G|=N_k|P|$.
The momentum vectors are then split into approximately $N_k/|P|$ momentum stars (with each star specified by the same data), where $|P|$ is the order of the crystal point group.
We discuss the detailed implementation of space-group adaptation further in section~\ref{sec:be}.

\subsection{Block encoding and data loading}
\label{sec:be}

We now determine the quantum resource cost of quantum phase estimation (QPE)~\cite{kitaev1995quantum} for computing eigenvalues of the BSE Hamiltonian $H_{\mathrm{BSE}}$.
We do not include the initial state preparation cost, although adiabatic state preparation is a natural strategy~\cite{aspuru2005simulated,albash2018adiabatic} for weakly- to moderately-correlated semiconductor excitonic states, starting from the orbital product states.
We use a fault-tolerant cost model of Toffoli count and logical qubit count.
In this cost model, the standard method is to use qubitization \cite{low2019qubitization,babbush2018encoding,berry2019qubitization,lee2021even}, where one constructs the walk operator $W=(2|0\rangle \langle 0| \otimes I - I)U$, where $U$ is a self-inverse block encoding of $H_\mathrm{BSE}$.
The eigenvalues of $W$ are $e^{\pm i \arccos (E/ \lambda)}$, where $E$ are the eigenvalues of $H_\mathrm{BSE}$ and $\lambda$ is the subnormalization of the block encoding.
As shown in Ref.~\cite{lee2021even}, to estimate an eigenvalue to accuracy $\epsilon_{\mathrm{QPE}}$, the total cost is
\begin{align}
  C_{\mathrm{QPE}}
  =
  \left\lceil \frac{\pi\lambda}{2\epsilon_{\mathrm{QPE}}}\right\rceil
  C_{\mathrm{walk}} .
\end{align}
Up to the small number of Toffolis associated with the reflection operation in $W$, $C_\mathrm{walk}$ is equivalent to the block-encoding cost, thus both $C_{\mathrm{walk}}$ and $\lambda$ are set by the block encoding method.

A standard block encoding strategy is to decompose the Hamiltonian into a linear combination of unitaries (LCU)~\cite{childs2012hamiltonian}, with $H = \sum_\mu c_\mu U_\mu$, where $U_\mu$ does not require an additional complicated synthesis, and with a total subnormalization $\lambda = \sum_\mu |c_\mu|$.
Examples of this approach include the sum of Paulis LCU in second quantized quantum simulation~\cite{babbush2018encoding}, where the coefficients are the integrals, e.g. $V_{pq,rs}$, or the tensor hypercontraction analog, where the coefficients are e.g. $\zeta_{\QQ,\mu\nu}^V$, and the unitaries are Givens rotations of Pauli operators~\cite{lee2021even,rubin2023faulttolerant}.

However, the BSE Hamiltonian, and especially its exchange part, has a different structure, that makes an analogous LCU summation (and in particular a summation over $\mathbf{Q}$) unfavourable.
Physically, treating an electron--hole pair as a single exciton (mathematically grouping the $ia$ and $jb$ indices in Eq.~\ref{eq:H_ov_ex}), the exchange Hamiltonian for one pair can be written conceptually as
\begin{equation} \label{eq:conceptual_H_ex}
  H_{ov}^{\mathrm{ex,proto}}
  =\sum_{\boldsymbol{Q}}V(\boldsymbol{Q})
  |\text{exciton},\boldsymbol{Q}\rangle
  \langle\text{exciton},\boldsymbol{Q}|.
\end{equation}
Here $|\text{exciton},\boldsymbol{Q}\rangle$ represents a Bloch exciton state with momentum $\boldsymbol{Q}$.
In the long-range asymptotic limit, $V(\boldsymbol{Q})$ is governed by the dipole--dipole kernel $\boldsymbol{Q}\boldsymbol{Q}/|\boldsymbol{Q}|^2-I/3$ (see details in Appendix~\ref{app:bse-norm}), therefore each $\boldsymbol{Q}$ has almost uniform $O(1)$ operator norm, while the full exchange Hamiltonian also has $O(1)$ operator norm.
Therefore, a naive LCU sum over $\boldsymbol{Q}$ introduces an unnecessary $O(N_k)$ higher subnormalization, even if each momentum block is encoded optimally.

We thus use a more natural LCU construction of $H_\mathrm{BSE}$.
More specifically, from Eq.~\ref{eq:H_BSE}, we have (i) a sum over 5 terms (templates), (ii) within each term, we have a sum over the quasiparticles (electrons, $e$, or holes $h$) and occupied/virtual combinations, and then finally (iii) a summation over the orbital labels, which yields the prototerms, which we can construct by direct synthesis.
For example, for $H_{oo}^{\mathrm{dir}}$, the prototerm is
\begin{align}
  H_{oo}^{\mathrm{dir,proto}} &= \sum_{\sigma \sigma'} \sum_{ijkl} W_{ki,lj}\;
  \ketbra{i\,\sigma}{k\,\sigma}\otimes\ketbra{j\,\sigma'}{l\,\sigma'}.
  \label{eq:proto_oo}
\end{align}
The prototerm is the same for any pair of holes.
Thus the summation in (i) is implemented by a PREPARE and standard SELECT LCU, but in (ii) PREPARE is followed by a SELECT which is implemented by a routing network which routes the selected hole registers $h,h'$ to a single synthesized prototerm.
The resulting Toffoli cost and subnormalization are given by
\begin{align}
  C_{\mathrm{walk}} &= C_0+C_{oo}^{\mathrm{dir}}+C_{vv}^{\mathrm{dir}}
  +C_{ov}^{\mathrm{ex}}+C_{ov}^{\mathrm{dir}}+C_{\mathrm{extra}},
  \label{eq:C_walk}\\
  \lambda &= \lambda_0+\lambda_{\mathrm{ex}}+\lambda_{\mathrm{dir}},
  \label{eq:lambda}\\
  \lambda_0 &= m\tilde{\lambda}_0,\qquad
  \lambda_{\mathrm{ex}} = m^2\tilde{\lambda}_{ov}^{\mathrm{ex}},
  \label{eq:lambda_fock_exchange}\\
  \lambda_{\mathrm{dir}}
  &= \binom{m}{2}\bigl(\tilde{\lambda}_{oo}^{\mathrm{dir}}
  +\tilde{\lambda}_{vv}^{\mathrm{dir}}\bigr)
  +m^2\tilde{\lambda}_{ov}^{\mathrm{dir}}.
  \label{eq:lambda_direct}
\end{align}
Here $C_0$ and $\lambda_0$ are the walk cost and subnormalization of the Fock term $H_0$.
$\lambda$ (without the tilde) denotes subnormalizations of the full $m$-exciton effective Hamiltonian and its components, while $\tilde{\lambda}$ denotes those of individual one- or two-body terms acting on one or two electrons or holes.
The cost $C_{\mathrm{extra}}$ contains extra subleading terms, such as preparing the LCU coefficients, register routing, momentum arithmetic, Fourier transforms, and reflection operator for the walk operator (note all such costs are included in our numerical resource estimates).
For gapped systems at fixed basis resolution and fitting accuracy, the prototerm subnormalizations $\tilde{\lambda}$ and, at fixed $m$, the total subnormalization $\lambda$, are theoretically expected to scale asymptotically as $O(1)$ (see Appendix~\ref{app:thc-scaling}).
In the numerical results for the Coulomb interaction below, we observe an effective scaling $\lambda \propto N^\alpha$ over the system sizes studied, with a small exponent $\alpha \ll 1$.
Since $m = O(1)$, for $N$ of interest, the total subnormalization is within a small factor of the optimal subnormalization of $H_\mathrm{BSE}$.

We now discuss the synthesis and subnormalization of the prototerms.
We note that the Fock term $H_0$ is diagonal in the orbital-energy basis, and the two-particle matrices $X\in\{W,V\}$ are block-diagonal due to translational symmetry.
We will use optimized direct synthesis methods based on the interferometer-based unitary synthesis in Sec III of~\cite{berry2024rapid}, a modified column-by-column isometry synthesis~\cite{iten2015quantum}, and QROAM-loaded rotations for diagonal operators, all adapted to block-diagonal matrices.

\begin{figure}[t]
\centering
\includegraphics[width=\columnwidth]{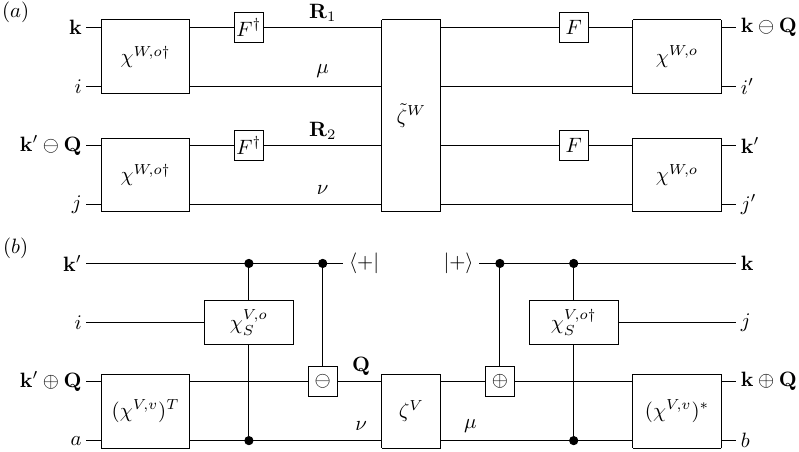}
\caption{\textbf{Block encodings of the direct and exchange term.}
(a) The direct $oo$ circuit, where $F$ is the Fourier transform on the momentum registers, $\chi^{W,o}$ is the block isometry of Eq.~\eqref{eq:be_bold_chi_main}, and $\tilde\zeta^W$ is the diagonal operator of Eq.~\eqref{eq:be_bold_zeta_R_main}.
The $vv$ and direct $ov$ circuits use the same structure.
(b) The exchange $ov$ circuit, where $\chi_S^{V,o}$ is the state preparation of Eq.~\eqref{eq:be_exchange_stateprep_main}, controlled by $\kk$ and the auxiliary index. }
\label{fig:thc-circuits}
\end{figure}

The Fock term can be synthesized easily and contributes little to the total cost, so we start with the two-body $oo$, $vv$, and $ov$ direct terms.
Since the procedure is similar for all three, we continue with the $oo$ example.
Using the factorized integrals, we rewrite the same prototerm defined in Eq.~\eqref{eq:proto_oo} as
\begin{widetext}
\begin{equation}
  H_{oo}^{\mathrm{dir,proto}}=\frac{1}{N_k}\sum_{\kk\kk'\QQ}\sum_{\mu\nu}
  \zeta^{W}_{\QQ,\mu\nu}
  \Bigl[\sum_{ii'}\chi^{W}_{i'(\kk\ominus\QQ),\mu}
  \chi^{W*}_{i\kk,\mu}\,
  \ketbra{i',\kk\ominus\QQ}{i,\kk}\Bigr]
  \otimes
  \Bigl[\sum_{jj'}\chi^{W}_{j'\kk',\nu}
  \chi^{W*}_{j(\kk'\ominus\QQ),\nu}\,
  \ketbra{j',\kk'}{j,\kk'\ominus\QQ}\Bigr]
  \label{eq:be_Hoo_proto}
\end{equation}
\end{widetext}
where we have suppressed the spin summation for simplicity.
(Technically, if we include spin, $H_{oo}^{\mathrm{dir,proto}} \rightarrow H_{oo}^{\mathrm{dir,proto}} \otimes I_\sigma$, where $I$ is an identity operator in the spin sector and does not increase the subnormalization, and the additional synthesis can be implemented with subleading costs).

Define the block-diagonal matrix
\begin{equation}
  \chi^{W,o}
  = \sum_{\kk i\mu}\chi^{W}_{i\kk,\mu}
  \ketbra{\kk i}{\kk\mu},
  \label{eq:be_bold_chi_main}
\end{equation}
$H_{oo}^{\mathrm{dir,proto}}$ can thus be expressed as
\begin{equation}
  H_{oo}^{\mathrm{dir,proto}}=\bigl(\chi^{W,o}\otimes\chi^{W,o}\bigr)
  Z_{oo}^{\mathrm{dir}}\bigl(\chi^{W,o}\otimes\chi^{W,o}\bigr)^{\dagger}
\end{equation}
with
\begin{equation}
  Z_{oo}^{\mathrm{dir}}=\frac{1}{N_k}\sum_{\kk\kk'\QQ}\sum_{\mu\nu}\zeta^{W}_{\QQ,\mu\nu}\,
  \ketbra{\kk\ominus\QQ,\mu}{\kk,\mu}\otimes
  \ketbra{\kk',\nu}{\kk'\ominus\QQ,\nu}
  \label{eq:be_Zoo_main}
\end{equation}
and our task is to encode the operators $\chi^{W,o}$ and $Z_{oo}^{\mathrm{dir}}$.

We first transform to real space where $Z_{oo}^{\mathrm{dir}}$ is diagonal, which means that loading the diagonal entries achieves the optimal subnormalization.
Define the Fourier-transformed central factor $\tilde\zeta^W_{\RR,\mu\nu}=N_k^{-1}\sum_{\QQ} e^{\ii\QQ\cdot\RR}\zeta^W_{\QQ,\mu\nu}$, together with the diagonal operator
\begin{equation}
  {\tilde\zeta}^{W}
  =\sum_{\RR_1\mu\RR_2\nu}
  \tilde\zeta^{W}_{\RR_1\ominus\RR_2,\mu\nu}
  \ketbra{\RR_1\mu\RR_2\nu}{\RR_1\mu\RR_2\nu}.
  \label{eq:be_bold_zeta_R_main}
\end{equation}
Then $Z_{oo}^{\mathrm{dir}}=(F\otimes F)\,\tilde\zeta^{W}(F\otimes F)^{\dagger}$, where $F\ket{\RR}=N_k^{-1/2}\sum_{\kk}e^{\ii\kk\cdot\RR}\ket{\kk}$ is the Fourier transform on a momentum register.
To encode $\tilde{\zeta}^{W}$, we compute $R=R_1\ominus R_2$ in place for addressing and use QROAM~\cite{berry2019qubitization,low2024trading} to load the elements $\tilde{\zeta}^{W}_{R,\mu\nu}/\|\tilde{\zeta}^{W}\|$ via the the angles $\theta_{R\mu\nu}=\arccos\bigl(\tilde{\zeta}^{W}_{R,\mu\nu} /\|\tilde{\zeta}^{W}\|\bigr)$ and a multiplexed rotation to transfer them into the block encoding~\cite{sunderhauf2024blockencoding} with subnormalization $\|\tilde{\zeta}^{W}\|$; the angle and the difference are subsequently uncomputed.

Using $\tilde{\zeta}^W$ and the Fourier transform circuit, we compose the block-encoding circuit shown in Figure~\ref{fig:thc-circuits}(a).
The matrix $\chi^{W,o}$ contains $N_k$ blocks of size $N_o\times\NTHC$.
We normalize all blocks by the common norm $\lVert\chi^{W,o}\rVert$, embed each block into an isometry, and synthesize the blocks together using the column-by-column method described in Appendix~\ref{app:block-isometry}, where the dominant cost is again the QROAM data-loading of rotation angles.
The $vv$ and direct $ov$ terms have the same structure, with the appropriate occupied or virtual factors and Fourier directions.
Note that there are shared circuit components; for example, the $oo$, $vv$, and direct $ov$ terms share the same ${\chi^{W,o}}, {\chi^{W,v}}, {{\tilde\zeta}^{W}}$.
The cost accounting for these shared components is summarized in Table~\ref{tab:primitives} of Appendix~\ref{app:primitives}.
The primitive direct subnormalizations are computed as
\begin{align}
  \tilde{\lambda}_{oo}^{\mathrm{dir}}
  &= \lVert\chi^{W,o}\rVert^4\lVert{\tilde\zeta}^{W}\rVert,
  \label{eq:be_lambda_ss_main}\\
  \tilde{\lambda}_{vv}^{\mathrm{dir}}
  &= \lVert\chi^{W,v}\rVert^4\lVert{\tilde\zeta}^{W}\rVert,\\
  \tilde{\lambda}_{ov}^{\mathrm{dir}}
  &= \lVert\chi^{W,o}\rVert^2
  \lVert\chi^{W,v}\rVert^2
  \lVert{\tilde\zeta}^{W}\rVert.
  \label{eq:be_lambda_dir_main}
\end{align}
Here $\lVert{\tilde\zeta}^{W}\rVert$ is the largest absolute real-space matrix element in Eq.~\ref{eq:be_bold_zeta_R_main}.

We next describe the synthesis cost and the subnormalization of the exchange $ov$ term.
With THC integrals, and temporarily neglecting spin, this takes the form
\begin{widetext}
\begin{equation}
  H_{ov}^{\text{ex,proto}}=\frac{1}{N_k}\sum_{\kk\kk'\QQ}\sum_{\mu\nu}
  \zeta^{V}_{\QQ,\mu\nu}
  \Bigl[\sum_{ij}\chi^{V}_{j\kk,\mu}\chi^{V*}_{i\kk',\nu}\,
  \ketbra{j,\kk}{i,\kk'}\Bigr]_{h}
  \otimes
  \Bigl[\sum_{ab}\chi^{V*}_{b(\kk\oplus\QQ),\mu}
  \chi^{V}_{a(\kk'\oplus\QQ),\nu}\,
  \ketbra{b,\kk\oplus\QQ}{a,\kk'\oplus\QQ}\Bigr]_{e}
\end{equation}
\end{widetext}

Factoring out the orbital factors as before gives
\begin{equation}
  H^{\text{ex,proto}}_{ov}=\bigl(\chi^{V,o}\!\otimes(\chi^{V,v})^{*}\bigr)
  Z^{\text{ex}}_{ov}\bigl(\chi^{V,o}\!\otimes(\chi^{V,v})^{*}\bigr)^{\dagger}
\end{equation}

with
\begin{equation}
  \begin{split}
  Z^{\text{ex}}_{ov}=\frac{1}{N_k}\sum_{\kk\kk'\QQ}\sum_{\mu\nu}
  \zeta^{V}_{\QQ,\mu\nu}\,
  &\ketbra{\kk,\mu}{\kk',\nu}_{h}\\
  &\otimes\,\ketbra{\kk\oplus\QQ,\mu}{\kk'\oplus\QQ,\nu}_{e}.
  \end{split}
\end{equation}
Two features distinguish this from $Z_{oo}^{\mathrm{dir}}$ of Eq.~\eqref{eq:be_Zoo_main}.
First, $\mu$ labels both kets and $\nu$ both bras, so $\zeta^{V}$ acts as a matrix from $\nu$ to $\mu$ rather than as a coefficient.
Second, the momenta enter as correlated electron--hole pairs, $(\kk,\kk\oplus\QQ)$ on the ket side and $(\kk',\kk'\oplus\QQ)$ on the bra side, sharing a common exciton momentum $\QQ$.
Both features are expressed by the two isometries,
\begin{equation}
  T=\frac{1}{\sqrt{N_k}}\sum_{\kk\QQ}
  \bigl(|\kk\rangle|\kk\oplus\QQ\rangle\bigr)\langle\QQ|,
  \qquad
  \Phi=\sum_{\mu}\bigl(|\mu\rangle|\mu\rangle\bigr)\langle\mu|,
\end{equation}
which build the correlated momentum pair and duplicate the auxiliary label respectively.
Defining the block-diagonal matrix,
\begin{equation}
  \zeta^{V}
  = \sum_{\QQ\mu\nu}\zeta^{V}_{\QQ,\mu\nu}
  \ketbra{\QQ\mu}{\QQ\nu},
  \label{eq:be_bold_zeta_Q_main}
\end{equation}
with dense square blocks of sizes $N_{\text{THC}} \times N_{\text{THC}}$, we have
\begin{equation}
  Z^{\text{ex}}_{ov}=(T\otimes\Phi)\,\zeta^{V}(T\otimes\Phi)^{\dagger},
\end{equation}

To block encode $\zeta^V$, we decompose each block into the $U D U^\dagger$ form, construct the eigenbasis transformation using the block unitary synthesis~\cite{berry2024rapid,rupprecht2026faster} detailed in Appendix~\ref{app:eigen} and encode the eigenvalues as a diagonal operator.
Similarly to the direct case, we encode $\chi^{V,v}$ as a block isometry.
For $\chi^{V,o}$, we find an improved Toffoli count and subnormalization by fusing $\chi^{V,o}$ with $\Phi$ (which consumes one of the $\mu$ copies in $\Phi$, leaving the other to act on by $\chi^{V,v}$).
The fused operator corresponds to
\begin{align}
  &\chi_S^{V,o}
  \coloneqq \sum_{\kk\mu}\ketbra{\kk\mu}{\kk\mu}
  \left[\left(\frac{1}{\gamma_o}\sum_i\chi^{V}_{i\kk,\mu}\bra{i} \right)+ \bra{\perp_{\kk,\mu}}\right],
  \label{eq:be_exchange_stateprep_main}\\
  &\gamma_o\coloneqq\max_{\kk,\mu}
  \left(\sum_i\left|\chi^{V}_{i\kk,\mu}\right|^2\right)^{1/2}.
  \label{eq:gamma_o}
\end{align}
where $\bra{\perp_{\kk,\mu}}$ is an orthogonal component that normalizes the state, which will be projected out in the block-encoding.
Compared with the block isometry encoding, this replaces the operator norm with $\gamma_o$, the largest column norm over $(\kk,\mu)$.
The overall circuit construction is shown in Fig.~\ref{fig:thc-circuits}(b).
Together, these factors give the primitive exchange subnormalization
\begin{equation}
  \tilde{\lambda}_{ov}^{\mathrm{ex}}
  \coloneqq 2\lVert\chi^{V,v}\rVert^2
  \lVert\zeta^V\rVert
  \gamma_o^2.
  \label{eq:be_lambda_exchange_main}
\end{equation}
where the additional factor of 2 now comes from restoring the treatment of spin.

The leading synthesis cost above comes from the use of QROM/QROAM data loading (Appendix~\ref{app:qroam}).
Here, one can use a Toffoli-ancilla tradeoff to change the Toffoli cost from $O(\mathrm{data})$ to $O(\sqrt{\mathrm{data}})$ using $O(\sqrt{\mathrm{data}})$ ancillas.
Using the cost formulae in Table~\ref{tab:primitives} of Appendix~\ref{app:primitives}, and taking the $N_{\mathrm{THC}}$ parameter assumed in earlier second quantized work~\cite{rubin2023faulttolerant} $N_{\mathrm{THC}}/N_{\mathrm{AO}}=8$ and a rotation bit-precision $b=32$, the Toffoli counts per walk query for the qubit- and Toffoli-optimized versions have leading orders
\begin{align}
  C_{\mathrm{walk}}
  &\approx
  \begin{cases}
  313\,N_k\,\NAO^2, & \text{Qubit-optimal},\\
  1268\,\sqrt{N_k}\,\NAO^{3/2}, & \text{Toffoli-optimal},
  \end{cases}
\end{align}
and the leading-order peak ancilla count is approximately
\begin{align}
  A_{\mathrm{walk}}
  &\approx
  \begin{cases}
  96+\left\lceil\log_2\!\left(64N_k\NAO^2\right)\right\rceil,
  & \text{Qubit-optimal},\\
  45\,\sqrt{N_k}\,\NAO, & \text{Toffoli-optimal}.
  \end{cases}
\end{align}

Space-group adaptation, introduced in Sec.~\ref{sec:integrals}, further reduces the data that must be loaded.
Here, this denotes exploiting crystal symmetries beyond the translations already incorporated throughout.
An element $g$ of the space group $G$ defined there maps Bloch orbitals at $\kk$ to orbitals at $g\kk$, atomic orbital components to (superposition of) other atomic orbital components, and maps interpolation points to symmetry-equivalent points.
We denote its unitary representation on the orbital index $(p,\kk)$ by $D_g^{\mathrm{orb}}$, which is $D_g^{o}$ or $D_g^{v}$ in the occupied or virtual sector, and its representation on the THC index $(\kk,\mu)$ by $D_g^{\mathrm{THC}}$.
The latter is a phased permutation: it permutes momenta and interpolation points and attaches a Bloch phase (Appendix~\ref{app:S}).
The THC fitting of Sec.~\ref{sec:thc-opt} makes the factors symmetric,
\begin{align}
  D_g^{\mathrm{orb}}\,\chi^X\,D_g^{\mathrm{THC}\dagger}&=\chi^X, \nonumber\\
  D_g^{\mathrm{THC}}\,\zeta^X\,D_g^{\mathrm{THC}\dagger}&=\zeta^X,
  \qquad g\in G .
  \label{eq:thc_symmetric}
\end{align}
Let $S^{\mathrm{orb}}$ and $S^{\mathrm{THC}}$ be unitaries that decompose the two representations into irreducible representations (irreps)~\cite{bradley2010mathematical}, $S^{t\dagger}D_g^{t}S^{t}=\bigoplus_\lambda D_\lambda(g)\otimes I_{n^t_\lambda}$ for $t\in\{\mathrm{orb},\mathrm{THC}\}$, where irrep $\lambda$ has dimension $m_\lambda$ and multiplicity $n^t_\lambda$.
By Schur's lemma, Eq.~\eqref{eq:thc_symmetric} makes the factors block diagonal in these bases,
\begin{align}
  \zeta^X_{\mathrm{sg}}
  &= S^{\mathrm{THC}\dagger}\zeta^X S^{\mathrm{THC}}
  =\bigoplus_{\lambda}
  I_{m_\lambda}\otimes\zeta^X_\lambda,
  \label{eq:be_zeta_sg_main}\\
  \chi^X_{\mathrm{sg}}
  &= S^{\mathrm{orb}\dagger}\chi^X S^{\mathrm{THC}}
  =\bigoplus_{\lambda}
  I_{m_\lambda}\otimes\chi^X_\lambda,
  \label{eq:be_chi_sg_main}
\end{align}
with $\zeta^X_\lambda$ of size $n^{\mathrm{THC}}_\lambda\times n^{\mathrm{THC}}_\lambda$ and $\chi^X_\lambda$ of size $n^{\mathrm{orb}}_\lambda\times n^{\mathrm{THC}}_\lambda$.

The original factors are recovered exactly as $\chi^X=S^{\mathrm{orb}}\chi^X_{\mathrm{sg}}S^{\mathrm{THC}\dagger}$ and $\zeta^X=S^{\mathrm{THC}}\zeta^X_{\mathrm{sg}}S^{\mathrm{THC}\dagger}$.
In the space-group-adapted block encoding, we therefore synthesize $\chi^X_{\mathrm{sg}}$ and $\zeta^X_{\mathrm{sg}}$ instead of the original factors wherever possible.
Specifically, we synthesize the block operators $\chi^{W,o}$, $\chi^{W,v}$, $\chi^{V,v}$ and $\zeta^V$ through their space-group-adapted forms, using the procedures above with blocks indexed by $\lambda$, and transform them with $S^{\mathrm{orb}}$ and $S^{\mathrm{THC}}$.
The identity factor $I_{m_\lambda}$ means that all $m_\lambda$ copies of a block share the same data, so only the blocks $\chi^X_\lambda$ and $\zeta^V_\lambda$ are loaded.
The remaining operators, $\tilde\zeta^W$ and $\chi^{V,o}_S$, do not take the form of a block-diagonal operator conjugated by $S^{\mathrm{THC}}$.
We therefore synthesize them in the original translation-only form.
Finally, $S^{\mathrm{orb}}$ is a unitary basis change that does not change the spectrum, so it need not be implemented (up to the initial state).
The cost of $S^{\mathrm{THC}}$ is subleading; its synthesis is described in Appendix~\ref{app:S}.

Now we analyze the qualitative reduction of cost by space-group adaptation.
Symmetry-related momenta form a momentum star, with one representative in the irreducible Brillouin zone.
A generic star contains $|P|$ momenta whose blocks share the same independent data, reducing the number of independently synthesized blocks from $N_k$ to approximately $N_k/|P|$.
High-symmetry points, lines, and planes can have smaller stars, but their fraction of a uniformly refined momentum mesh vanishes in the thermodynamic limit.
The leading qubit-optimal and Toffoli-optimal costs therefore decrease by factors of approximately $|P|$ and $\sqrt{|P|}$, respectively.

\begin{table*}[t]
\centering\small
\setlength{\tabcolsep}{5pt}
\renewcommand{\arraystretch}{1.18}
\begin{tabular*}{\textwidth}{@{\extracolsep{\fill}}lcccc@{}}
  \toprule
  \multicolumn{5}{@{}l}{\textbf{(a) Quantum Algorithms}}\\
  \addlinespace[3pt]
  \textbf{Method}
  & \multicolumn{3}{c}{\textbf{Toffoli count}}
  & \multicolumn{1}{c@{}}{\textbf{Logical qubits}}\\
  \cmidrule(lr){2-4}\cmidrule(l){5-5}
  & {\footnotesize\shortstack{Subnormalization\\$\lambda$}}
  & {\footnotesize\shortstack{Block-encoding\\$C_{\mathrm{walk}}$}}
  & \shortstack{\textbf{Total}\\$\lambda C_{\mathrm{walk}}$}
  & \\
  \midrule
  \textbf{Quantum BSE T-optimal} & {\footnotesize $N^\alpha$} & {\footnotesize $N^{1/2}$} & $N^{1/2+\alpha}$ & $N^{1/2}$\\
  \textbf{Quantum BSE Q-optimal} & {\footnotesize $N^\alpha$} & {\footnotesize $N$} & $N^{1+\alpha}$ & $\log N$\\
  Second quantization, THC
  & {\footnotesize $N^2$} & {\footnotesize $N$} & $N^3$ & $N$\\
  First quantization
  & {\footnotesize $N^2$} & {\footnotesize $N$} & $N^3$ & $N$\\
  \midrule
  \multicolumn{5}{@{}l}{\textbf{(b) Classical Algorithms}}\\
  \addlinespace[3pt]
  \textbf{Method}
  & \multicolumn{3}{c}{\textbf{Time}}
  & \multicolumn{1}{c@{}}{\textbf{Space}}\\
  \midrule
  Classical BSE
  & \multicolumn{3}{c}{$N^{2m}$} & $N^{2m-1}$\\
  \bottomrule
\end{tabular*}
\caption{\textbf{Resource scaling for crystalline systems with periodic boundary conditions.}
The system size is varied through the number of translationally equivalent primitive cells $N_k$, keeping the orbital basis per primitive cell, THC rank ratio $c_{\mathrm{IP}}$, and electron--hole pair number $m$ fixed.
Thus $N=N_k\NAO\propto N_k$, and the tabulated powers describe scaling with the number of primitive cells.
(a) Quantum resources at fixed QPE precision.
Under a fixed QPE target error, the total Toffoli scaling is proportional to the product of the subnormalization $\lambda$ and block-encoding cost $C_{\mathrm{walk}}$.
For quantum BSE, $\lambda\propto N^\alpha$, with theoretical asymptotic exponent $\alpha=0$ (Appendices~\ref{app:bse-norm} and~\ref{app:thc-scaling}); finite-size numerical estimates instead give a small effective exponent $\alpha\approx0.2$ (Fig.~\ref{fig:optimized-channel-scaling}).
(b) Classical BSE, exploiting translational symmetry at fixed total crystal momentum.
Classical exponents depend explicitly on $m$, whereas quantum BSE exponents are independent of fixed $m$.
Logarithmic factors are omitted for the full-Hamiltonian quantum methods.}
\label{tab:simple_scaling}
\end{table*}

\subsection{THC optimization}
\label{sec:thc-opt}

We initialize THC through the interpolable separable density fitting form~\cite{lu2015compression,hu2017interpolative}, where the index $\mu$ corresponds to real-space grid points at $\boldsymbol{r}_{\mu}$, selected by a pivoted Cholesky operation on the orbital product matrix evaluated on a predefined grid in the cell.
Then the $\boldsymbol{\chi}^{X}$ tensors are simply the orbital values on the grids: $\chi^X_{p\boldsymbol{k},\mu}=\phi_{p\boldsymbol{k}}(\boldsymbol{r}_{\mu})$.
Once the $\boldsymbol{\chi}^{X}$ tensors are determined, the $\boldsymbol{\zeta}^{X}$ tensors can be obtained by a by least-squares fit~\cite{parrish2012tensor}, with a regularization to balance between accuracy and sub-normalization~\cite{lee2021even}.

The least-squares loss function is conceptually written as
\begin{equation} \label{eq:loss}
  L=\log(\text{relative ERI error})+\alpha(\text{subnormlization}),
\end{equation}
where the relative ERI error is given by
\begin{equation} \label{eq:rel_error}
  \text{relative ERI error}=\frac{\Vert X_{\text{fit}}-X\Vert_{2}}{\Vert X\Vert_{2}}
\end{equation}
where $X\in\{W,V\}$ as above, and $\alpha$ serves as an externally-controlled hyperparameter to balance accuracy and sub-normalization.
In practice, the loss function form is slightly adapted for better empirical performance.

For each system, we sweep over a few different values of $\alpha$ to obtain a set of THC decompositions with different accuracies and subnormalizations.
For the bare interaction, we assess the fitting accuracy using the second-order Moller-Plesset (MP2)~\cite{moller1934note} energy error per cell, following Ref.~\cite{rubin2023faulttolerant}.
The diamond data show an overall increase in MP2 error with the relative fitting error of the bare $(ov|vo)$ block.
We use this relation to calibrate relative-error targets, which are then applied to the $(ov|W|vo)$ projection of the fitted full screened interaction, as detailed in Appendix~\ref{app:fit_THC}.

To perform the space-group-adapted THC fitting, one additionally requires that the orbitals can be labelled by the additional space-group labels.
A sufficient (but not necessary) condition that the fitted THC form has the full $A_{1}$ symmetry, is that the $\boldsymbol{\chi}^{X}$ and {$\boldsymbol{\zeta}^{X}$} tensors both have full $A_{1}$ symmetry.
To use this, we explicitly symmetrize the input $\boldsymbol{\chi}^{X}$ tensor by
\[
  \boldsymbol{\chi}^{X}\rightarrow\frac{1}{|G|}\sum_{g\in G}D_{g}^{\text{orb}}\boldsymbol{\chi}^{X}D_{g}^{\text{THC},\dagger},
\]
where $D_{g}^{\text{orb}}$ is one of $D_{g}^{o},D_{g}^{v}$, the representation of the element $g$ in the orbital space, and similarly for $D_g^{\text{THC}}$.
The symmetry of $\boldsymbol{\zeta}^{X}$, obtained from the least-squares solution, is automatically guaranteed by this structure.
The same loss function is then used for the optimization.
In practice, we find that the assumption of fully symmetric $\boldsymbol{\chi}^{X}$ and $\boldsymbol{\zeta}^{X}$ does not lead to a significant loss of accuracy.

\subsection{Summary of asymptotic speedups}
\label{sec:asymptotic}

We now describe the asymptotic speedups achievable by quasiparticle quantum simulation of the multi-exciton BSE Hamiltonian, versus first- and second-quantized simulation of the full Hamiltonian, as well as standard classical simulation.
For the quantum algorithms, we report the Toffoli and qubit cost of $\lambda C_\text{walk}$, while for the classical algorithm, we give the scaling assuming BSE is solved iteratively by dense matrix-vector multiplications~\cite{deslippe2012berkeleygw,sangalli2019many}.
We consider increasing the number of translationally equivalent primitive cells $N_k$ in a supercell with periodic boundary conditions, keeping the primitive-cell structure, orbital basis per cell $\NAO$, and THC rank ratio $c_{\mathrm{IP}}$ fixed.
The size parameter is therefore $N=N_k\NAO\propto N_k$.
In Table~\ref{tab:simple_scaling} we show simplified scalings at fixed electron--hole pair number $m$, retaining the explicit $m$ dependence of the classical exponents.
Detailed costs of the block-encoding primitives, including their dependence on block dimensions and rotation precision, are given in Appendix~\ref{app:primitives}.

The subnormalization is theoretically expected to scale asymptotically as $\lambda=O(1)$ at fixed excitation number and accuracy (Appendix~\ref{app:thc-scaling}), giving $O(N^{1/2})$ and $O(N)$ Toffoli costs for the Toffoli- and qubit-optimized algorithms, respectively.
Retaining the effective numerical scaling $\lambda \propto N^\alpha$, quasiparticle simulation has a cost of $O(N^{1/2+\alpha})$ Toffolis and $O(N^{1/2})$ qubits (Toffoli-optimized), and $O(N^{1 + \alpha})$ Toffolis and $O(\log N)$ qubits (qubit efficient), with a small numerical exponent $\alpha \ll 1$ over the system sizes studied (see next section).
For classical BSE, exploiting translational symmetry at fixed total crystal momentum reduces the cost of a matrix--vector multiplication from $O(N^{2m+2})$ to $O(N^{2m})$ and the storage from $O(N^{2m})$ to $O(N^{2m-1})$.
For triexcitonic states ($m=3$), this gives $O(N^6)$ time and $O(N^5)$ space.
Approximating $\alpha \approx 0$, we achieve up to a 12th power time speedup (Toffoli-optimized) and an exponential savings in space (qubit-optimized algorithm) over the default classical algorithm.

First- and second-quantized algorithms using LCU block encodings for full Hamiltonian simulation require $\widetilde{O}(N)$ Toffolis to apply terms across the electron or orbital registers \cite{berry2024quantum,rubin2023faulttolerant}.
The associated subnormalization is also expected to scale as $\widetilde{O}(N^2)$: in second quantization, THC sums magnitudes of effective interaction coefficients and exhibits approximately quadratic growth in materials benchmarks \cite{lee2021even,rubin2023faulttolerant}, while in first quantization it depends explicitly on the $O(N^2)$ electron pairs \cite{berry2024quantum}.
For excitations, resolving a fixed absolute excitation energy further requires computing both ground- and excited-state total energies to fixed additive accuracy, unlike per-cell accuracy which permits total-energy errors that grow with $N$.
Thus at fixed excitation-energy precision, this yields an overall Toffoli cost of $\widetilde{O}(N^3)$, while storing the state requires $O(N)$ or $O(N \log N)$ qubits.
Assuming $\alpha \approx 0$, the quasiparticle quantum simulation thus achieves a 6th power speedup in time (Toffoli-optimized) and an exponential savings in space (qubit-optimized) over full Hamiltonian simulation (although, of course, the problem being simulated is not identical).
As discussed above, this speedup comes from reducing the block-encoding cost, where the only $N$ scaling comes from data-loading, as well as the favourable subnormalization, which physically arises from the $O(1)$ spectral width of $H_{\mathrm{BSE}}$.

{We note briefly that there are very recent full-Hamiltonian simulation algorithms that improve on this LCU scaling.
Trotter-based methods with tighter error bounds and cheaper Trotter steps~\cite{su2021nearly,low2023complexity,berry2026fmm} can reach $N^{5/3+o(1)}$ gates when the numbers of electrons and orbitals both scale as $N$, and sum-of-squares spectral amplification~\cite{king2026quantum,low2025spectral,dutkiewicz2026spectral} reduces the subnormalization, by a factor of $O(N^{1/2})$ in first quantization.
We do not use these methods as baselines, since comparable end-to-end resource estimates for the materials and system sizes considered here are largely unavailable.
We note that their asymptotic costs remain well above the $O(N^{1/2+\alpha})$ Toffoli scaling of the quasiparticle simulation.}

\begin{table}[bp]
\centering
\begin{ruledtabular}
\begin{tabular}{lrrr}
  System & Atoms & Active electrons & $N_{\mathrm{AO}}$ \\
  \hline
  Diamond & 2 & 8 & 26 \\
  MgO     & 2 & 16 & 27 \\
  BN      & 2 & 8 & 26 \\
  AlN     & 4 & 16 & 52 \\
  CdSe (zinc blende) & 2 & 18 & 38 \\
  MoSe$_2$ (monolayer) & 3 & 26 & 52 \\
\end{tabular}
\end{ruledtabular}
\caption{System information for resource estimates considered in this work.
All counts are per unit cell; $N_{\mathrm{AO}}$ is the number of atomic-orbital basis functions.
Diamond, MgO, BN, and AlN use the GTH-cc-pVDZ basis~\cite{ye2022correlation}, while CdSe and MoSe$_2$ use the GTH-DZVP-MOLOPT-SR basis~\cite{vandevondele2007gaussian}.}
\label{tab:system-information}
\end{table}

\begin{figure*}[htbp]
\centering
\includegraphics[width=\textwidth]{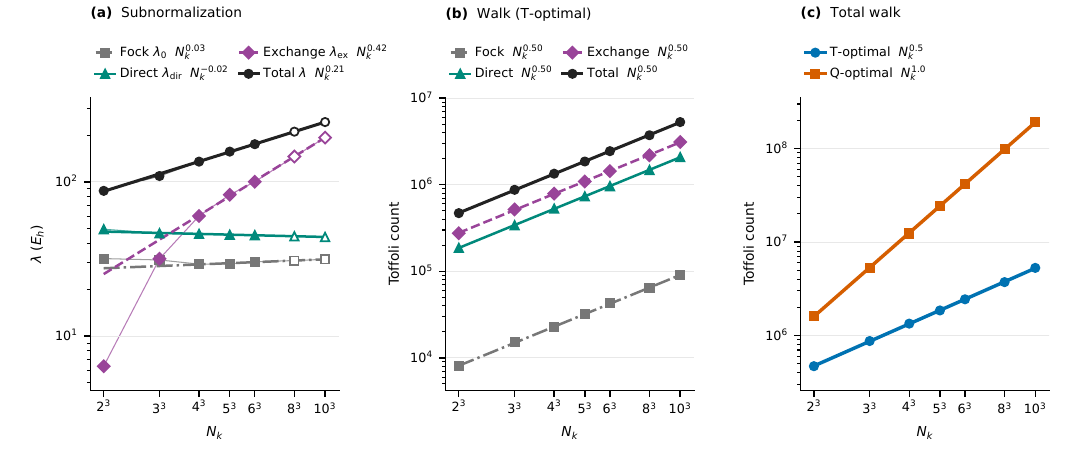}
\caption{Subnormalizations and walk costs for triexcitonic BSE eigenvalue estimation in diamond without using space-group adaptation, using a THC fitting criterion of approximately $0.1\,\mathrm{m}E_h$ MP2 energy error per unit cell.
(a) Fock, direct, exchange, and total subnormalizations, including the $m=3$ excitation-order factors defined in Eqs.~\eqref{eq:lambda_fock_exchange} and \eqref{eq:lambda_direct}.
Thin lines connect successive points, scaling fits are obtained from the last three data points, and open markers denote extrapolated data.
(b) Toffoli-optimized walk components and their sum.
(c) Total walk costs for both T-optimal and Q-optimal compilation; the legend gives the theoretical leading powers $N_k^{0.5}$ and $N_k^{1.0}$.}
\label{fig:optimized-channel-scaling}
\end{figure*}

\section{Resource estimates for multi-exciton simulation in semiconductors}

We now present logical resource estimates for the BSE simulation of multi-excitonic states.
As above, we choose the $m=3$ triexcitonic state setting as one which balances physical relevance and classical difficulty.
Classical preprocessing of the data is performed using PySCF~\cite{sun2018pyscf,sun2020pyscf}, and we also use its implementation of periodic GW (specifically, G$_0$W$_0$)~\cite{zhu2021allelectron}.
The calculations used the GTH-cc-pVDZ basis of Ye and Berkelbach~\cite{ye2022correlation} for diamond, MgO, BN, and AlN, and use the GTH-DZVP-MOLOPT-SR basis~\cite{vandevondele2007gaussian} for zinc-blende CdSe and monolayer MoSe$_2$.

The materials were mainly chosen for benchmarking value rather than for a direct relevance to multi-excitonic physics.
Diamond, BN, and AlN have previously been studied in the context of first and second quantized simulation resource estimation~\cite{rubin2023faulttolerant,berry2024quantum}.
MgO is a material with a strongly bound $m=1$ exciton~\cite{fuchs2008efficient}, and CdSe is a material studied in the context of multi-exciton generation in nanocrystals and engineered nanostructures~\cite{xiao2012carrier,noh2025highly}.
Monolayer MoSe$_2$ belongs to the two-dimensional transition-metal dichalcogenides, whose strongly bound excitons and biexcitons have been studied extensively~\cite{steinhoff2018biexciton,torche2021biexcitons}.
However, even for CdSe, which we treat here as a periodic bulk zinc-blende crystal, the focus of the current work is on resource estimation as a foundation for future simulations of the multi-exciton generation process.
Table~\ref{tab:system-information} summarizes the numbers of atoms, active electrons, and basis functions per unit cell for the various semiconductors.

We first examine the subnormalization and walk costs, then compare the resulting resource scaling with standard full Hamiltonian quantum simulation at the same target accuracy (a typical electronic spectroscopy computational accuracy of $0.1\,\mathrm{eV}$).

\begin{figure*}[htbp]
\centering
\includegraphics[width=0.95\textwidth]{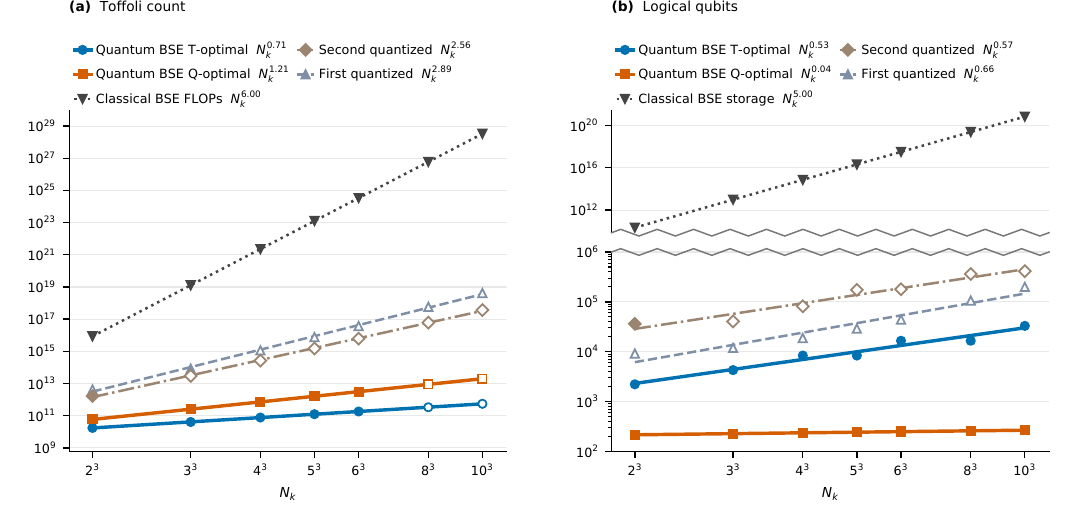}
\caption{Quantum and classical resource estimates for diamond.
The BSE calculations use $m=3$, a THC fitting tolerance of $0.1\,\mathrm{m}E_h$ MP2 error per unit cell, and no space-group adaptation.
All quantum methods target $0.1\,\mathrm{eV}$ QPE error per eigenvalue estimate: an excitation energy for BSE and a total supercell energy for the full-Hamiltonian methods.
Constant overheads for combining ground- and excited-state estimates are omitted.
(a) Quantum Toffoli counts and classical BSE FLOPs.
(b) Logical qubits and classical BSE storage, counted in floating-point values.
Open markers denote extrapolated data.}
\label{fig:optimized-literature-comparison}
\end{figure*}

\subsection{Resource scaling of walk cost and subnormalization}
\label{sec:results-scaling}

We first examine the resources required for each step of the algorithm and their dependence on $N_k$, using diamond as an example.
Here, we do not use space-group adaptation.
To be consistent with Ref.~\cite{rubin2023faulttolerant}, we fit the THC factors so that the estimated ground-state MP2 error is approximately $0.1\,\mathrm{m}E_h$ per unit cell or less.
The detailed parameters are described in Appendix~\ref{app:fit_THC}.
Figure~\ref{fig:optimized-channel-scaling} shows the subnormalizations in panel (a), the Toffoli-optimized counts for different walk components in (b), and the total walk costs in (c).

In panel (a), we see that the Fock and direct contributions to the subnormalization are nearly independent of system size, whereas exchange grows as a function of the mesh size.
Its growth rate nevertheless decreases for larger $N_k$, consistent with the arguments for $\sim O(1)$ asymptotic scaling given in Appendices~\ref{app:bse-norm} and~\ref{app:thc-scaling}.
Fitting to the last three measured meshes gives $N_k^{0.42}$ for exchange and $N_k^{0.21}$ for the total, and the total subnormalization increases only from $87.5$ to $175.7\,E_h$ as $N_k$ increases $27$-fold, from $8$ to $216$.
The open markers show extrapolations to $N_k=512$ and $1000$.

Panel (b) shows that the direct and exchange terms dominate the Toffoli-optimized walk cost, while the Fock contribution is smaller.
(Here the resource estimates allow for general occupied and virtual Fock blocks $f^o$ and $f^v$; using their diagonal orbital-energy representation would give a small savings).
The walk costs are evaluated from the circuit-cost formulae, which allows us to directly extract their scaling.
Their leading system-size dependence is $N_k^{0.5}$ under Toffoli optimization and $N_k^{1.0}$ under qubit optimization, as derived in Sec.~\ref{sec:asymptotic}.
Panel (c) displays these two behaviors.
The slow growth of the subnormalization therefore provides only a modest additional system-size dependence in the total phase-estimation cost.

\subsection{Comparison to first- and second-quantized simulations}
\label{sec:results-diamond}

We compare quantum BSE with first- and second-quantized full-Hamiltonian simulations at a QPE error of $0.1\,\mathrm{eV}\simeq3.68\,\mathrm{m}E_h$ per eigenvalue estimate.
BSE directly estimates excitation energies, whereas the full-Hamiltonian methods require differences between independently resolved total supercell energies.
The second-quantized reference's original tolerance of $1\,\mathrm{kcal\,mol^{-1}}$ per cell~\cite{rubin2023faulttolerant} must therefore be converted to fixed supercell accuracy, introducing an additional factor of $N_k$ in the QPE cost at the same numerical tolerance; the first-quantized reference already uses total-energy accuracy~\cite{berry2024quantum}.
Figure~\ref{fig:optimized-literature-comparison} shows the resulting estimates, retaining the preceding subsection's THC criterion of $0.1\,\mathrm{m}E_h$ MP2 error per cell and using no space-group adaptation.
Details of the reference estimates are given in Appendix~\ref{app:baselines}.

We show the Toffoli counts in panel (a) of Fig.~\ref{fig:optimized-literature-comparison}.
The BSE Toffoli counts scale as $N_k^{0.71}$ and $N_k^{1.21}$ under Toffoli and qubit optimization, respectively.
These exponents are consistent with the theoretical $N_k^{1/2+\alpha}$ and $N_k^{1+\alpha}$ scalings in Table~\ref{tab:simple_scaling}, with the weak subnormalization growth $\alpha\approx0.2$ observed above.
The full-Hamiltonian estimates scale as $N_k^{2.56}$ in second quantization and $N_k^{2.89}$ in first quantization over the plotted range.
This is below their theoretical asymptotic cubic dependence because at these sizes, the lower-order contributions still dominate.

Due to the difference in scaling, the resource advantage of quantum simulation of the BSE problem versus the full Hamiltonian problem increases with system size.
At $N_k=1000$, the Toffoli-optimized BSE Toffoli count is $5.5\times10^{11}$, compared with $3.7\times10^{17}$ and $4.2\times10^{18}$ for second and first quantization, respectively, i.e. we achieve a $10^6-10^7 \times$ reduction.
The corresponding qubit-optimized BSE Toffoli count is $2.0\times10^{13}$.

Panel (b) shows the logical qubit requirements.
The Toffoli-optimized BSE qubit count scales approximately as $N_k^{0.53}$, close to the theoretical square-root dependence, while the qubit-optimized count grows only from $215$ to $266$, consistent with logarithmic scaling.
The full-Hamiltonian qubit counts have effective exponents $0.57$ and $0.66$ over this range, below the asymptotic linear dependence in Table~\ref{tab:simple_scaling}.
At $N_k=1000$, the BSE formulation requires $3.3\times10^4$ or $266$ qubits under the two optimization choices, compared with $4.1\times10^5$ and $2.0\times10^5$ for second and first quantization.

Finally, in both panels, we give a classical reference cost for a linear matrix-vector solve, assuming 100 iterations and dense matrix-vector multiplication, with cost $100N_o^3N_v^5N_k^6$ FLOPs.
At $N_k=1000$, the classical calculation requires $3.30\times10^{28}$ FLOPs and $6.8\times10^{20}$ words of memory, both far exceeding any practical computational resource today.

\begin{figure}[t]
\centering
\includegraphics[width=\columnwidth]{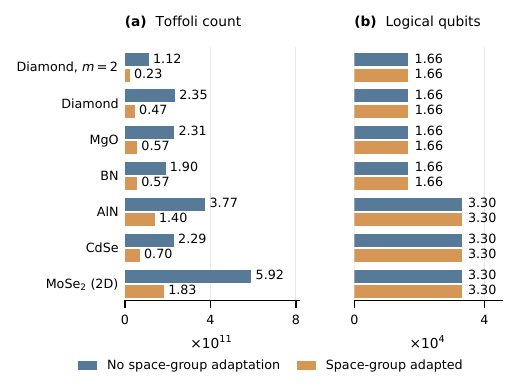}
\caption{Toffoli-optimized multi-exciton BSE resources at $N_k=1000$ ($32\times32=1024$ for monolayer MoSe$_2$), with a target QPE error of $0.1\,\mathrm{eV}$ in the excitation energy and a THC fitting tolerance below $0.005\,\mathrm{eV}$ MP2 error per active electron.
All cases use $m=3$ unless otherwise labelled.
(a) Toffoli counts without (blue) and with (orange) space-group adaptation.
Diamond uses subnormalizations from space-group-adapted fits; the other materials retain those from fits without space-group adaptation.
(b) Logical-qubit counts, which agree between the two settings at the displayed precision.
Input estimates and subnormalization extrapolations are detailed in Appendix~\ref{app:resource-tables}.}
\label{fig:optimized-exciton-comparison}
\end{figure}

\subsection{Optimized resource estimates for BSE calculations}
\label{sec:results-dynamics}

We now obtain our best quantum resource estimates by further optimizing the computational settings for multi-exciton BSE calculations targeting $0.1\,\mathrm{eV}$ resolution in the excitation energy on the $10\times10\times10$ $k$-point mesh ($32\times32$ for monolayer MoSe$_2$).
We focus on minimizing the Toffoli count.
Since we are working with a quasiparticle theory and few-body Hamiltonians, it is reasonable to consider the THC error as incurred per electron.
We thus adjust the conservative THC fitting criterion used above to achieve an {MP2} error of below $0.005\,\mathrm{eV}$ per active electron, and use space-group adaptation.
The corresponding relative-integral-error criterion for the screened interaction is described in Appendix~\ref{app:fit_THC}.
Details of the extrapolated inputs and error conventions are given in Appendix~\ref{app:resource-tables}.

Diamond has $48$ point-group operations, giving an ideal Toffoli-optimized reduction of approximately $\sqrt{48}\simeq6.9$.
At fixed THC ranks, space-group adaptation reduces the walk cost by a factor of $4.0$: the direct central tensor $\tilde\zeta^W$ and exchange state preparation remain in the physical basis, while the $S^{\mathrm{THC}}$ transforms add overhead.
Space-group-adapted fitting also yields a modestly smaller subnormalization at comparable relative ERI error: for $m=3$ and $N_k=1000$, the extrapolated value decreases from $157.3$ to $126.7\,E_h$, which we attribute to the improved optimization landscape.
Together, these effects reduce the Toffoli count by a factor of $5.0$, to $4.7\times10^{10}$, while the logical-qubit count remains $1.7\times10^4$ to the displayed precision.
Following previous estimates \cite{lee2021even, rubin2023faulttolerant, gidney2025how}, assuming each Toffoli gate requires around 25 $\mu$s, we obtain a total runtime of about 14 days (as opposed to an estimated $10^5$--$10^6$ years with existing full Hamiltonian simulation protocols).

For MgO, BN, AlN, CdSe, and MoSe$_2$,\footnote{For monolayer MoSe$_2$, the symmetry group is the layer group, with point group $D_{3h}$ (12 operations).} we combine the space-group-adapted walk costs with the subnormalizations from the fits without space-group adaptation.
These fixed-subnormalization estimates yield Toffoli reductions of $2.7$--$4.0$, as shown in Fig.~\ref{fig:optimized-exciton-comparison}; detailed values are provided in Appendix~\ref{app:resource-tables}.

Without space-group adaptation, diamond, MgO, and BN have similar costs, between $1.9\times10^{11}$ and $2.3\times10^{11}$ Toffolis.
These systems have similar basis sizes of $26$--$27$ atomic orbitals per unit cell in the GTH-cc-pVDZ basis (Table~\ref{tab:system-information}).
AlN has $52$ orbitals per unit cell in the same basis and correspondingly higher requirements, at $3.8\times10^{11}$ Toffolis and $3.3\times10^4$ qubits, compared with $1.7\times10^4$ qubits for the other three materials.
Zinc-blende CdSe has $38$ basis functions and $18$ active electrons per primitive cell, so the full simulation is of 38,000 orbitals.
The triexcitonic calculation then requires approximately $2.3\times10^{11}$ Toffolis and $3.3\times10^4$ logical qubits.
Monolayer MoSe$_2$ has $52$ basis functions and $26$ active electrons per cell, so on the $32\times32$ mesh the full simulation is of 53,248 orbitals.
It requires $5.9\times10^{11}$ Toffolis and $3.3\times10^4$ logical qubits, the largest Toffoli count among the materials, mainly due to its larger subnormalization of $168\,E_h$.
Since its direct term uses the bare Coulomb interaction, this subnormalization is an upper bound.

\section{Conclusions}

We have presented a detailed quasiparticle quantum simulation framework for multi-excitonic states within the Bethe-Salpeter equation.
Our analysis of scaling advantage versus standard full Hamiltonian quantum simulation reveals up to 6th power asymptotic speedup in Toffoli gates and up to an exponential reduction in qubit count for quantum phase estimation.
We further described an explicit first-quantized, orbital-based, block-encoding compilation and carried out detailed resource estimates for the classically challenging, but physically relevant, problem of tri-excitonic eigenvalue estimation across a range of semiconductor models.
For the diamond benchmark system, at a realistic Brillouin zone sampling of $10\times 10\times 10$, we achieve a $10^6-10^7$ reduction in Toffoli gate count versus standard full Hamiltonian simulation, reducing the estimated quantum run time from $10^5$--$10^6$ years to roughly two weeks.

Overall, this work illustrates the importance of existing classical heuristic frameworks in the era of fault-tolerant quantum simulation.
The applicability of the heuristic approximations is to the same physical problem domains as studied in the classical simulations; however, quantum algorithms allow the possibility of extending the problem domains in practice, for example, as pursued here in the description of multi-excitonic as opposed to single-excitonic solutions of the Bethe-Salpeter equation.
In the context of practical quantum advantage, further restriction of the simulation problem, for example, in the cases here to low-energy multi-excitonic states and dynamics, may introduce further structure that can be incorporated into classical heuristics.
The open question in those settings is whether such structure may also be advantageously incorporated into the quantum algorithm.
However, our work supports the broad thesis that existing classical heuristic ideas contain essential information that should be utilized in simulations of the physical world, and that designing quantum algorithms in the regimes where classical heuristics work~\cite{chen2025framework} will be an essential direction in the further development of the field.

\bibliography{references.bib}

\clearpage
\onecolumngrid
\section*{Supplementary Material}

\appendix

\section{Primitives used for block-encoding and their costs}
\label{app:primitives}

The BSE block-encoding mainly uses block unitary synthesis, block isometry synthesis, block state preparation, and diagonal matrix synthesis.
We give the Toffoli and ancilla costs of each primitive used to construct the proto-terms in the BSE block-encoding.
Table~\ref{tab:primitives} illustrates their composition and leading costs for the translation-only construction, where each primitive acts on blocks of equal size.
Space-group adaptation produces blocks of varying sizes; the same constructions apply using the primitives described below, so we omit a separate summary table.
The additional space-group transformation is treated in Appendix~\ref{app:S}.

\begin{table}[H]
\centering
{\footnotesize
\renewcommand{\arraystretch}{1.12}
\setlength{\tabcolsep}{3pt}
\begin{tabular*}{\textwidth}{@{\extracolsep{\fill}}lccccc@{}}
  \specialrule{0.8pt}{0pt}{0pt}
  \multicolumn{6}{@{}l}{\textbf{(a) Factors in each term}}\\
  \specialrule{0.8pt}{2pt}{2pt}
  \textbf{Term} & $\chi^{o}$ & $\chi^{v}$ & $\chi^{o}_{S}$ & $\zeta^{V}$ & $\zeta^{W}$\\
  \specialrule{0.8pt}{2pt}{2pt}
  Fock & --- & --- & --- & --- & --- \\
  $oo$ direct & 4 & --- & --- & --- & 1 \\
  $vv$ direct & --- & 4 & --- & --- & reused \\
  $ov$ direct & reused & reused & --- & --- & reused \\
  $ov$ exchange & --- & 2 & 2 & 1 & --- \\
  \specialrule{0.8pt}{2pt}{2pt}
  Walk operator & 4 & 6 & 2 & 1 & 1 \\
  \specialrule{0.8pt}{0pt}{0pt}
\end{tabular*}}

\vspace{8pt}

{\footnotesize
\renewcommand{\arraystretch}{1.12}
\setlength{\tabcolsep}{2.7pt}
\begin{tabular*}{\textwidth}{@{\extracolsep{\fill}}lcccccccc@{}}
  \specialrule{0.8pt}{0pt}{0pt}
  \multicolumn{9}{@{}l}{\textbf{(b) Primitives used by each factor}}\\
  \specialrule{0.8pt}{2pt}{2pt}
  \multirow{2}{*}{\textbf{Part}}
  & \multicolumn{2}{c}{\textbf{Block isometry}}
  & \multicolumn{2}{c}{\textbf{Block  state prep.}}
  & \multicolumn{2}{c}{\textbf{Block unitary}}
  & \multicolumn{2}{c}{\textbf{Diagonal}}\\
  \cmidrule[0.6pt](lr){2-3}\cmidrule[0.6pt](lr){4-5}\cmidrule[0.6pt](lr){6-7}\cmidrule[0.6pt](lr){8-9}
  & count & (blocks $B$, rows $N$, cols $M$) & count & (blocks $B$, entries $N$)
  & count & (blocks $B$, rows \& cols $N$) & count & entries $N$\\
  \specialrule{0.8pt}{2pt}{2pt}
  $\chi^{o}$
  & 1 & $(N_k,\NTHC+N_o,N_o)$ & --- & --- & --- & --- & --- & --- \\
  $\chi^{v}$
  & 1 & $(N_k,\NTHC+N_v,N_v)$ & --- & --- & --- & --- & --- & --- \\
  $\chi^{o}_{S}$
  & --- & --- & 1 & $(N_k\NTHC,N_o)$ & --- & --- & --- & --- \\
  $\zeta^{V}$
  & --- & --- & --- & --- & 2 & $(N_k,\NTHC)$ & 1 & $N_k\NTHC$ \\
  $\zeta^{W}$
  & --- & --- & --- & --- & --- & --- & 1 & $N_k\NTHC^{2}$ \\
  \addlinespace[2pt]
  \shortstack[l]{$f^{o},\,f^{v}$\\(Fock blocks)}
  & --- & --- & --- & --- & 4 & $(N_k,N_o),\ (N_k,N_v)$
  & 2 & $N_kN_o,\ N_kN_v$ \\
  \specialrule{0.8pt}{2pt}{2pt}
  Walk operator
  & 10 & --- & 2 & --- & 6 & --- & 4 & --- \\
  \specialrule{0.8pt}{0pt}{0pt}
\end{tabular*}}

\vspace{8pt}

{\scriptsize
\renewcommand{\arraystretch}{1.20}
\setlength{\tabcolsep}{2.5pt}
\begin{tabular*}{\textwidth}{@{\extracolsep{\fill}}llcccc@{}}
  \specialrule{0.8pt}{0pt}{0pt}
  \multicolumn{6}{@{}l}{\textbf{(c) Leading resource cost of each primitive}}\\
  \specialrule{0.8pt}{2pt}{2pt}
  \multirow{2}{*}{\textbf{Primitive}}
  & \multirow{2}{*}{\textbf{Operator}}
  & \multicolumn{2}{c}{\textbf{Q-optimal}}
  & \multicolumn{2}{c}{\textbf{T-optimal}}\\
  \cmidrule[0.6pt](lr){3-4}\cmidrule[0.6pt](lr){5-6}
  & & \textbf{Toffolis} & \textbf{Ancillas} & \textbf{Toffolis} & \textbf{Ancillas}\\
  \specialrule{0.8pt}{2pt}{2pt}
  Block isometry synthesis
  & $\bigoplus_{k=1}^{B}V_k,\quad V_k\in\mathbb C^{N\times M}$
  & $2MNB$ & $3b+\log_2(NB)$
  & $7\sqrt bM\sqrt{NB}$ & $\sqrt{bNB}$ \\
  \addlinespace[3pt]
  Block state preparation
  & $\bigoplus_{k=1}^{B}S_k,\quad S_k\ket{0}=\ket{\psi_k}\in\mathbb C^{N}$
  & $2NB$ & $3b+\log_2(NB)$
  & $3\sqrt{bNB}+2bN$ & $\sqrt{2bNB}$ \\
  \addlinespace[3pt]
  Block unitary synthesis
  & $\bigoplus_{k=1}^{B}U_k,\quad U_k\in\mathbb C^{N\times N}$
  & $BN^{2}$ & $3b+\log_2(NB)$
  & $2\sqrt{bB}\,N^{3/2}$ & $\sqrt{bNB}$ \\
  \addlinespace[3pt]
  Diagonal (real entries)
  & $\sum_{n=1}^{N}d_n\ketbra{n}{n},\quad d_n\in\mathbb R$
  & $N$ & $3b+\log_2 N$
  & $2\big(\sqrt b+1\big)\sqrt{N}$ & $\sqrt{bN}$ \\
  \specialrule{0.8pt}{0pt}{0pt}
\end{tabular*}}
\caption{\textbf{Primitive composition and leading costs for the translation-only block-encoding.}
(a) gives the factors whose product forms each term of the BSE block-encoding, the entries counting how often each factor appears; (b) gives the primitives each factor is synthesized from; and (c) gives the Toffoli and ancilla cost of each primitive for equal-sized blocks.
The interaction label on $\chi^o,\chi^v$ is suppressed in this cost table: direct terms use $\chi^{W,o},\chi^{W,v}$, exchange uses $\chi^{V,v}$ and the state-preparation operator $\chi_S^{V,o}$, and the separately fitted $W$ and $V$ factors need not coincide.
Multiplying (a) by (b) yields the ``walk operator'' row of (b).
Each shared part is counted once, so the $oo$ direct term reuses the parts of the direct terms and adds no cost of its own.
The Fock term draws on no shared part, being the eigendecomposition of the occupied and virtual Fock blocks $f^{o}$ and $f^{v}$, and is listed in (b) so that the totals stay column sums.
Here $B$ is the block count, $N$ the matrix dimension, $M\leq N$ the number of isometry columns, and $b$ the angle precision.
For block isometries, $N$ is the padded row count.
The Q-optimal and T-optimal variants are defined in Appendix~\ref{app:qroam} and derived at the end of each primitive's subsection.
Numerical prefactors are rounded to integers and subleading terms omitted; unequal blocks arising from space-group adaptation are handled by the subset constructions below.}
\label{tab:primitives}
\end{table}

\subsection{QROM and QROAM data-loading}
\label{app:qroam}

All primitives will consist of a sequence of coherent data-loading of rotation angles, which are fed into multi-controlled rotation gates.
Our primary data-loading primitive is quantum read-only memory (QROM) and its improved version QROAM~\cite{berry2019qubitization}.
Given a table of $L$ entries $f(j)\in\{0,1\}^b$, QROM implements the coherent map $\ket{j}\ket{0}^{\otimes b} \longmapsto \ket{j}\ket{f(j)}$.
Ref.~\cite{babbush2018encoding} introduced unary-iteration QROM, which uses exactly $L-1$ Toffolis.
Ref.~\cite{low2024trading} introduced the Select--Swap network, enabling a Toffoli--ancilla trade-off that reduces the forward cost to $O(\sqrt{bL})$.
Ref.~\cite{berry2019qubitization} introduced QROAM, which uses measurement and classical feedforward to uncompute the data registers at cost $O(\sqrt{L})$ independent of the word size $b$.
In what follows, QROAM uses unary-iteration QROM as a subroutine, and all lookup costs include one external control.

In our applications, the set of valid addresses is often a subset $\mathcal S\subseteq\{0,\ldots,2^n-1\}$ (e.g., addresses distributed across padded blocks).
We therefore require a lookup that acts as the identity outside $\mathcal S$, namely the subset-restricted QROM
\begin{equation}
  \sum_{j\in\mathcal S}\ketbra{j}{j}\otimes X^{f(j)}
  +\sum_{j\notin\mathcal S}\ketbra{j}{j}\otimes I,
  \label{eq:qroam_safe}
\end{equation}
where $X^{f(j)}$ applies $X$ to each target qubit for which $f(j)$ has bit value $1$.
Relative to the usual unary-iteration formulation, (i) the address set need not be a contiguous interval such as $[0,L-1]$, and (ii) addresses outside the set are explicitly assigned the all-zero word, ensuring that they leave the output unchanged.
For a contiguous interval, this corresponds to ``restricted partial unary iteration'' as studied in Ref.~\cite{rupprecht2026sparse}; here we use the same idea for an arbitrary subset.

We construct subset-restricted QROM by pruning the binary-tree traversal used in unary iteration.
Leaves represent addresses, and internal nodes represent prefixes shared by the addresses beneath them.
Starting from the external control, the circuit checks successive address bits while descending the tree: each internal node uses one Toffoli to select its first child and a CNOT to switch to the other child.
At a leaf, CNOTs write the selected data word.
The circuit then returns up the tree, clearing temporary controls using measurements and Clifford corrections.
For subset restriction, we skip branches containing no address in $\mathcal S$ but keep all address checks on the remaining paths, so that any $j\notin\mathcal S$ leaves the output register unchanged.
The Toffoli cost is the number of internal nodes remaining after pruning.
For a fixed most-significant-bit-first traversal order,
\begin{equation}
  C_{\operatorname{QROM}}(\mathcal S,n)
  =\sum_{r=1}^{n}
  \left|\left\{
  \left\lfloor\frac{j}{2^r}\right\rfloor:j\in\mathcal S
  \right\}\right|.
  \label{eq:qrom_subset_cost}
\end{equation}
The circuit uses $n$ clean ancillas, excluding the address register, the output register, and the external control.
For $\mathcal S=\{0,\ldots,L-1\}$, each term equals $\lceil L/2^r\rceil$, giving $C_{\operatorname{QROM}}([0,L-1],n)=L-1+n-\operatorname{popcount}(L-1)$, where $\operatorname{popcount}$ counts ones in the binary representation, recovering the contiguous-interval formula in Ref.~\cite{rupprecht2026sparse}.

To obtain a subset-restricted QROAM, choose a power of two $\lambda\leq2^n$ and split the address as $j=q\lambda+r$ with quotient $q=\lfloor j/\lambda\rfloor$ and remainder $r=j\bmod\lambda$.
Define the quotient subset $\mathcal S_\lambda=\{\lfloor j/\lambda\rfloor:j\in\mathcal S\}$.
A quotient lookup loads a block of $\lambda$ candidate entries into $\lambda$ registers of $b$ qubits each (entries outside $\mathcal S$ are loaded as all zeros).
A swap tree controlled by the remainder moves the selected entry into the output register.
The swap network has $\lambda-1$ register swaps, each costing $b$ Toffolis, so the total forward cost is
\begin{equation}
  C_{\operatorname{QROAM}}(\mathcal S,n,b,\lambda)
  =C_{\operatorname{QROM}}(\mathcal S_\lambda,n-\log_2\lambda)
  +b(\lambda-1).
  \label{eq:qroam_subset_comp}
\end{equation}
The forward step uses $b(\lambda-1)+n-\log_2\lambda$ clean workspace qubits.
As in Appendix~B of Ref.~\cite{berry2019qubitization}, we may measure and release the unused data registers, storing their outcomes so that the induced phases can be corrected during the final uncomputation; Eq.~\eqref{eq:qroam_subset_comp} assumes this delayed correction.

For the QROAM uncomputation (Appendix~C of Ref.~\cite{berry2019qubitization}), the output word is measured in the $X$ basis and the outcomes are used to build a classical sign-fixup table.
Choose an independent power of two $2\leq\lambda'\leq2^n$.
The circuit (Fig.~6 of Ref.~\cite{berry2019qubitization}) encodes the remainder in a one-hot register, uses a quotient lookup to apply the required phase, and clears the one-hot register using measurements and Clifford corrections.
Only the quotient lookup requires the external control.
Preparing the one-hot register costs $\lambda'-2$ Toffolis.
For erasing a lookup result supported on $\mathcal S$, the cost is
\begin{equation}
  C_{\operatorname{QROAM}^{-1}}(\mathcal S,n,\lambda')
  =C_{\operatorname{QROM}}(\mathcal S_{\lambda'},n-\log_2\lambda')
  +\lambda'-2.
  \label{eq:qroam_subset_uncomp}
\end{equation}
Uncomputation uses $\lambda'+n-\log_2\lambda'$ clean workspace qubits; the measured output can be reset and reused for this workspace.
If phase corrections from unused forward registers were delayed, they must be included in the fixup table.
It is sufficient to include every address in any block loaded during the forward step.
When $\lambda'\geq\lambda$, these additional addresses do not change $\mathcal S_{\lambda'}$, so Eq.~\eqref{eq:qroam_subset_uncomp} applies as written; otherwise, the quotient subset must be enlarged to include their corrections.

Finally, subset restriction also covers multiple address registers: we concatenate their bits into one address and take $\mathcal S$ to be the Cartesian product of the valid ranges.
For QROAM, the concatenated high-order bits can be handled by subset-restricted QROM, and the remaining low-order bits select within the loaded block via the swap network.

The trade-off parameters are chosen in one of two ways, which define the Toffoli-optimized (T-optimal) and qubit-optimized (Q-optimal) variants of the primitives below and in Table~\ref{tab:primitives}(c).
Consider a lookup of $L$ words of $b$ bits.
In the T-optimal variant, $\lambda\simeq\sqrt{L/b}$ minimizes Eq.~\eqref{eq:qroam_subset_comp}, giving approximately $2\sqrt{bL}$ Toffolis with $\sqrt{bL}$ workspace qubits, and $\lambda'\simeq\sqrt L$ minimizes Eq.~\eqref{eq:qroam_subset_uncomp}, giving approximately $2\sqrt L$ Toffolis.
In the Q-optimal variant, the forward lookup uses $\lambda=1$, i.e.\ plain QROM, which costs approximately $L$ Toffolis and needs no data registers beyond the $b$-bit output.
The uncomputation parameter $\lambda'$ is not fixed to a small value such as $1$ or $2$: it is chosen to minimize the Toffoli count subject to not raising the peak qubit count.
The uncomputation reuses the measured output register and the released address controls, so every power of two with $\lambda'-\log_2\lambda'\leq b$ fits within the workspace of the forward step, and the optimum is the largest of them ($\lambda'=32$ for $b=32$).
The uncomputation then costs approximately $L/\lambda'$ Toffolis, a negligible correction to the forward lookup.
When a loaded angle is applied by addition into the phase-gradient register~\cite{gidney2018halving,sanders2020compilation}, the Q-optimal peak consists of the $b$-bit angle word, the $b$-qubit phase-gradient register, the $b-1$ carry qubits of the adder, and the address controls, i.e.\ approximately $3b+\log_2L$ qubits.

\subsection{Multiplexed rotations and diagonal matrix synthesis}

Throughout the construction of primitives, data-loading is mainly used to supply rotation angles to standard multiplexed single-qubit rotations~\cite{mottonen2004quantum,shende2006synthesis}
\begin{equation}
  U_{\mathrm{mux}}\;=\;\sum_{j=0}^{N-1} \ket{j}\bra{j} \otimes R_z(\theta_j),
\end{equation}
where the address register stores $j$ and the target is rotated by an angle $\theta_j$ specified by classical data.
One can convert to $R_x$ and $R_y$ rotations by Clifford gates.
In our implementation, the angles are stored to $b$ bits and are loaded coherently with subset-restricted QROAM on a set $\mathcal S$ of addresses that require a nontrivial rotation; after applying the rotation, the angle word is uncomputed.
For $n=\lceil\log_2 N\rceil$ address qubits and $\lambda'\geq\lambda$, the Toffoli cost is
\begin{equation}
  \begin{aligned}
  C_{\mathrm{mux}}
  ={}&C_{\operatorname{QROAM}}(\mathcal S,n,b,\lambda)+(b-2)\\
  &+C_{\operatorname{QROAM}^{-1}}(\mathcal S,n,\lambda').
  \end{aligned}
  \label{eq:mux_rotation_cost}
\end{equation}
where the explicit $(b-2)$ term accounts for applying the $b$-bit rotation once the angle has been loaded.

Several steps in the block-encoding reduce to applying diagonal operators (e.g., the diagonal real-space central factor).
A single multiplexed $R_y$ rotation is directly a block-encoding of a real diagonal matrix $D=\sum_{j=0}^{N-1}d_j\ketbra{j}{j}$, by treating the target qubit of the rotation as the ancilla and setting $\bra{0} R_y(2 \theta_j)\ket{0} = d_j/\alpha$.
To satisfy the self-inverse requirement, we append a single $Z$ gate since $(R_y(2\theta) Z)^\dagger = ZR_y(-2\theta)= R_y(2\theta)Z$.
Choose a normalization constant $\alpha\geq\max_j|d_j|$ with $\alpha>0$ so that each entry can be realized as a rotation amplitude.
Load $\theta_j=\arccos(d_j/\alpha)$ to $b$ bits, apply $R_y(2\theta_j)Z$ to a signal qubit, and uncompute the angle.
The signal-$\ket{0}$ block is $D/\alpha$ up to angle rounding.
Let $\mathcal S$ be the set of basis states $j$ such that $d_j\neq\alpha$, for $n=\lceil\log_2N\rceil$ and $\lambda'\geq\lambda$, the Toffoli cost is
\begin{equation}
  \begin{aligned}
  C_{\mathrm{diag}}
  ={}&C_{\operatorname{QROAM}}(\mathcal S,n,b,\lambda)+(b-2)\\
  &+C_{\operatorname{QROAM}^{-1}}(\mathcal S,n,\lambda').
  \end{aligned}
  \label{eq:diagonal_synthesis_cost}
\end{equation}
Optimized lookups give the leading cost $2(\sqrt b+1)\sqrt{|\mathcal S|}$ and workspace $\sqrt{b|\mathcal S|}$, recovering $2(\sqrt b+1)\sqrt N$ and $\sqrt{bN}$ in the dense case $\mathcal S=\{0,\ldots,N-1\}$.

For a dense diagonal with $N$ entries, the T-optimal variant loads the angles with $\lambda\simeq\sqrt{N/b}$ and erases them with $\lambda'\simeq\sqrt N$, which gives $2\sqrt{bN}+2\sqrt N=2(\sqrt b+1)\sqrt N$ Toffolis and $\sqrt{bN}$ workspace qubits.
The Q-optimal variant loads them with $\lambda=1$, at $N$ Toffolis, and erases them with the optimized $\lambda'$ of Appendix~\ref{app:qroam}, at $N/\lambda'$ Toffolis.
Its leading cost is therefore $N$ Toffolis with $3b+\log_2N$ qubits.
These are the diagonal entries of Table~\ref{tab:primitives}(c).

\subsection{Block-unitary synthesis}
\label{app:eigen}

The central $\zeta^V$ matrix in the exchange term is synthesized by performing an eigendecomposition and synthesizing the unitary eigentransformation directly.
We extend Refs.~\cite{berry2024rapid,rupprecht2026faster} to the synthesis of a block-diagonal unitary with unequal block sizes using subset-restricted QROAM.
Consider $B$ unitary blocks of size $N_i\times N_i$, each embedded into an $R_i\times R_i$ matrix with $N_i\leq R_i$.
We assume $N_i$ and $R_i$ are even (an odd dimension can be handled by padding with one identity row and column); all remaining rows and columns act as the identity.
Order the blocks by size, $N_0\geq N_1\geq\cdots\geq N_{B-1}$.
Block $i$ begins at address $s_i=\sum_{j<i}R_j$ in an $n=\lceil\log_2\sum_i R_i\rceil$-qubit address register.
Any unused basis states beyond the embedded blocks also act as the identity.
This formalism covers all unitary blocks in our block-encoding: for the eigenbasis transformation of $\boldsymbol{\zeta}^{V}$ under translation symmetry, $B=N_k$, $N_i=N_{\mathrm{THC}}$, and $R_i=2^{\lceil\log_2 N_{\mathrm{THC}}\rceil}$, whereas in the space-group-adapted case, $B$ is the number of irreducible representations and the unitaries act on the corresponding copy blocks with $N_i=R_i$.

\begin{figure}[t]
\centering
\begin{minipage}{0.82\textwidth}
\centering
\textbf{(a)}\par\vspace{0.25em}
\includegraphics[width=0.95\linewidth]{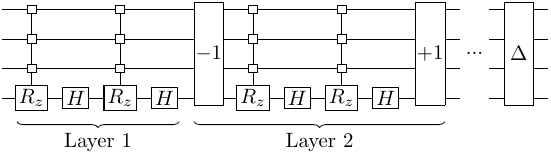}
\end{minipage}

\vspace{0.8em}

\begin{minipage}{\textwidth}
\centering
\textbf{(b)}\par\vspace{0.25em}
\includegraphics[width=\linewidth]{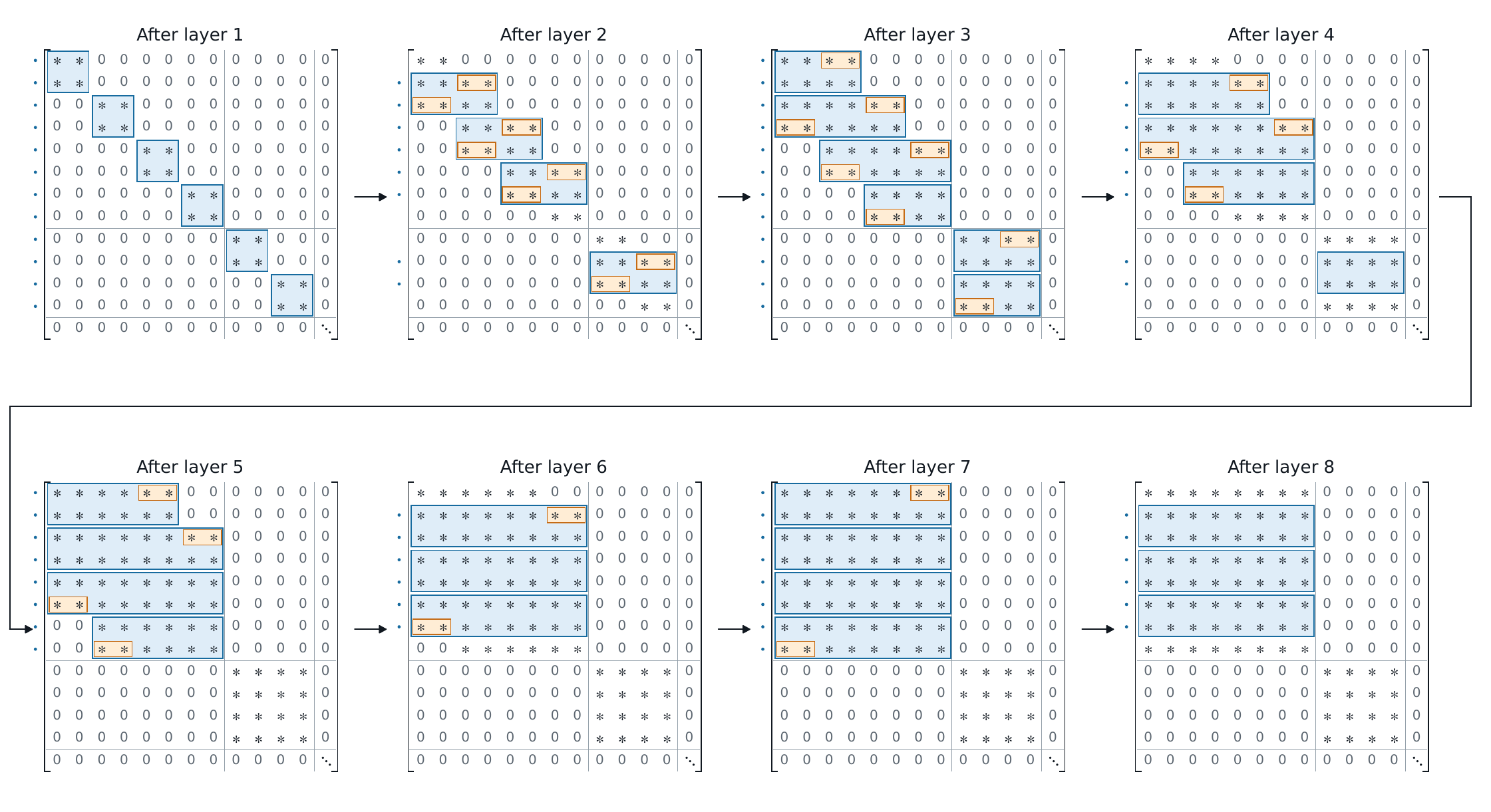}
\end{minipage}
\caption{\textbf{Block-unitary synthesis.}
(a) The circuit that synthesizes the block-unitary, composed of alternating layers followed by a diagonal phase layer $\Delta$.
White squares indicate multiplexed rotations; $\pm 1$ denotes increment/decrement.
(b) Growth of the synthesized blocks under the applications of each layer (final phase layer omitted).
Blue boxes mark entries updated by each layer, and orange boxes mark entries that were previously zero and become non-zero.}
\label{fig:block-unitary}
\end{figure}

The construction applies a sequence of layers that couple adjacent basis states.
Successive layers alternate between the two pairing patterns \begingroup
\setlength{\arraycolsep}{3pt}
\begin{equation}
  \begin{pmatrix}
  \ast & \ast &      &      &      &      &      &      \\
  \ast & \ast &      &      &      &      &      &      \\
  &      & \ast & \ast &      &      &      &      \\
  &      & \ast & \ast &      &      &      &      \\
  &      &      &      & \ast & \ast &      &      \\
  &      &      &      & \ast & \ast &      &      \\
  &      &      &      &      &      & \ast & \ast \\
  &      &      &      &      &      & \ast & \ast
  \end{pmatrix},
  \qquad
  \begin{pmatrix}
  \ast &      &      &      &      &      &      &      \\
  & \ast & \ast &      &      &      &      &      \\
  & \ast & \ast &      &      &      &      &      \\
  &      &      & \ast & \ast &      &      &      \\
  &      &      & \ast & \ast &      &      &      \\
  &      &      &      &      & \ast & \ast &      \\
  &      &      &      &      & \ast & \ast &      \\
  &      &      &      &      &      &      & \ast
  \end{pmatrix}.
  \label{eq:block-unitary-layer-patterns}
\end{equation}
\endgroup Here asterisks denote potentially nonzero entries and blanks denote zeros.
Each $2\times2$ unitary acting on a selected pair is decomposed into two phase rotations and two Hadamard gates, followed by diagonal phases.
The intermediate diagonal phases are pushed into subsequent layers, leaving a single diagonal phase layer at the end.
An ``odd'' layer (the right-hand pattern) is implemented by decrementing the address, applying the pair rotations, and incrementing the address.
Figure~\ref{fig:block-unitary} shows the circuit and how successive layers grow the synthesized blocks.

At each layer, the two $b$-bit angles for the $2\times2$ decomposition are loaded jointly via subset-restricted QROAM with trade-off parameter $\lambda$ and applied using addition into a phase-gradient state~\cite{sanders2020compilation}.
After application, the angle registers and any unused QROAM words are measured in the $X$ basis and reused; the resulting known signs are tracked classically and folded into later angle data and the final phase layer.
With this bookkeeping, each layer effectively costs one forward QROAM.

The Toffoli count is determined by the lookup subsets used in each subset-QROAM call.
At layer $\ell$, each unfinished block contributes $N_i/2$ lookup addresses, while blocks completed by layer $\ell$ contribute none, giving
\begin{equation}
  \mathcal S_\ell
  =
  \bigcup_{i:\,N_i>\ell}
  \left\{
  \frac{s_i}{2}+j:
  0\leq j<\frac{N_i}{2}
  \right\},
  \qquad
  |\mathcal S_\ell|
  =
  \frac12\sum_{i:\,N_i>\ell}N_i.
  \label{eq:block-unitary-angle-subset}
\end{equation}
When $R_i=N_i$, each subset is a contiguous prefix; padding introduces gaps, which subset-QROAM handles directly.

The final diagonal phase layer uses a subset $\mathcal S_{\mathrm{phase}}$ that contains all unitary-block addresses together with any padding or unused addresses that may acquire signs from QROAM measurements.
Because QROAM loads records in groups, a measurement sign can affect the $2\lambda$ addresses sharing the same upper bits (i.e., those obtained by varying the last $\log_2(2\lambda)$ bits).
We therefore include each such group and the address immediately after it (to account for odd-layer shifts), yielding a phase subset satisfying $\sum_i N_i\leq|\mathcal S_{\mathrm{phase}}|\leq 2^n$.
The final lookup and its erasure are performed only once, so they contribute only a small fraction of the total synthesis cost.

Assume $n,b\geq2$ and that a phase-gradient state is available.
Fix a power of two $1\leq\lambda\leq2^{n-1}$ for all forward lookups, including the final phase lookup.
Choose an independent power of two $\max\{2,\lambda\}\leq\lambda'\leq2^n$ for the final sign fixup.
The constraint $\lambda'\geq\lambda$ ensures that any additional addresses affected by measurements in the final phase lookup fall within the same $\lambda'$ quotient blocks already included in $\mathcal S_{\mathrm{phase}}$.
The Toffoli count is
\begin{equation}
  \begin{aligned}
  C_{\mathrm{block}}
  ={}&
  \sum_{\ell=0}^{N_0-1}
  \left[
  C_{\operatorname{QROAM}}(\mathcal S_\ell,n-1,2b,\lambda)
  +2(b-2)
  \right]
  +N_0(n-2)
  \\
  &+
  C_{\operatorname{QROAM}}(\mathcal S_{\mathrm{phase}},n,b,\lambda)
  +(b-2)
  \\
  &+
  C_{\operatorname{QROAM}^{-1}}(\mathcal S_{\mathrm{phase}},n,\lambda').
  \end{aligned}
  \label{eq:block-unitary-cost}
\end{equation}
The summand accounts for one angle lookup and two rotations per layer.
Each shift costs $n-2$ Toffolis, contributing $N_0(n-2)$.
The remaining terms apply the final phases and uncompute the lookup registers, including measurement-induced sign corrections.
Ancillas are reused throughout; phase-gradient-state preparation is counted separately.

For $B$ blocks of equal size $N$, the synthesis has $N$ layers, and each layer looks up the angles of $BN/2$ rotation pairs.
In the T-optimal variant, the two $b$-bit angles of a pair are loaded jointly as one $2b$-bit word, as in Eq.~\eqref{eq:block-unitary-cost}.
With $\lambda\simeq\sqrt{BN/(4b)}$, each layer costs approximately $2\sqrt{2b\cdot BN/2}=2\sqrt{bBN}$ Toffolis and uses $\sqrt{bBN}$ workspace qubits, so the $N$ layers cost $2\sqrt{bB}\,N^{3/2}$ Toffolis.
In the Q-optimal variant, a joint word would occupy $2b$ data qubits, so the two angles are instead loaded by two separate $\lambda=1$ lookups into a single $b$-bit register.
Each layer then costs $2\cdot BN/2=BN$ Toffolis, and the $N$ layers cost $BN^2$ Toffolis with $3b+\log_2(NB)$ qubits.
The final phase layer and its sign fixup, with $\lambda'$ optimized as in Appendix~\ref{app:qroam}, add only $O(BN)$ Toffolis.
These are the block-unitary entries of Table~\ref{tab:primitives}(c).

Recent advances in unitary synthesis could further reduce the walk cost.
For a single $N\times N$ unitary, recent constructions lower the $N^{3/2}$ scaling to $O(N^{4/3}\log^{2/3}N)$~\cite{tan2025unitary} and $\widetilde O(N^{5/4})$~\cite{yuan2026general} T gates, and to $\widetilde O(N)$ for near-Clifford unitaries~\cite{fang2026nearclifford}.
Since block-unitary synthesis dominates the $\zeta^V$ encoding, these could reduce the walk cost further.
However, their constant factors have not been worked out and may outweigh the modest asymptotic gain at the block sizes relevant here, so we leave them as potential future improvements.

\subsection{Block-isometry synthesis}
\label{app:block-isometry}

The occupied and virtual factors $\chi^{X,o}$ and $\chi^{X,v}$ contain blocks that have $N_{\mathrm{THC}}$ columns but only $N_o$ or $N_v$ rows.
Because the column-row ratio is large, embedding them into unitaries is expensive, motivating an isometry synthesis primitive, i.e. a unitary where only the first few columns need to be specified and the rest can be arbitrary.
We synthesize these columns by a column-elimination circuit and then reverse it.
This subsection is a blockwise extension of the column-by-column construction of Ref.~\cite{iten2015quantum} to unequal block sizes, with subset-restricted QROAM providing the phase and rotation angles shared across all active blocks.

Consider $B$ blocks $V_i\in\mathbb C^{N_i\times M_i}$ satisfying $V_i^\dagger V_i=I_{M_i}$ and $1\leq M_i\leq N_i$, ordered so that $N_0\geq\cdots\geq N_{B-1}$.
Pad block $i$ to $R_i=2^{p_i}$ rows, where $p_i=\lceil\log_2N_i\rceil$, and embed it at offset $s_i=\sum_{j<i}R_j$ in an $n=\lceil\log_2\sum_iR_i\rceil$-qubit register.
Row $x$ of block $i$ has global address $s_i+x$.
Since the padded sizes are decreasing powers of two, each block begins at a multiple of its own size; equivalently, the lowest $p_i$ address bits are exactly the local row bits.
This alignment lets us apply the same binary-tree rotation pattern in every block without mixing rows from different blocks.

We construct a column-elimination circuit $U^\dagger$ such that, for each block $i$ and each prescribed column $0\leq c<M_i$,
\begin{equation}
  U^\dagger
  \sum_{x=0}^{N_i-1}(V_i)_{xc}\ket{s_i+x}
  =\ket{s_i+c}.
  \qquad
  \label{eq:block_isometry_annihilation}
\end{equation}
Reversing this circuit gives the desired synthesis $U$, with input $\ket{s_i+c}$ selecting column $c$ of block $i$.
Although only these columns are specified, the resulting gate sequence fixes a full unitary completion on each padded block; addresses outside the embedded blocks are left unchanged by the corrected circuit.

The elimination proceeds column by column.
For a given column $c$, we (i) apply a diagonal phase layer to make the active entries real and nonnegative, and then (ii) apply a conditioned binary tree of real rotations to concentrate the remaining amplitude into the single target row $s_i+c$ in every block with $c<M_i$.
At tree level $t$, we rotate the qubit $n-t$ and condition on the lower $t$ bits matching those of $c$.
Because these lower bits are fixed by earlier levels, each enabled rotation combines the two surviving amplitudes in its selected pair, stores the norm in the member whose target-bit value matches $c$, and zeros the other.
We omit rotations that intersect already cleared rows or have zero angle; a zero entry does not by itself remove the operation if the nonzero entry lies in the ``wrong'' member of the pair (in which case a swap is still required).
Rotations act on all uncleared columns, so the phase and rotation tables used for later columns must be computed from the ideal elimination sequence.
Once a block has no remaining columns, it receives no further reference gates.
In particular, column $c$ requires at most $\max_{i:c<M_i}p_i$ rotation levels, so a large block with few columns can finish early, reducing the depth of later trees.

\begin{figure}[t]
\centering
\begin{minipage}{0.82\textwidth}
\centering
\textbf{(a)}\par\vspace{0.25em}
\includegraphics[width=0.95\linewidth]{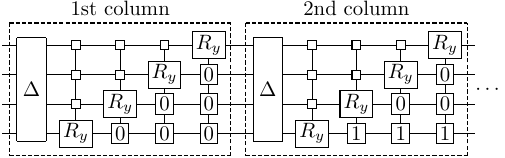}
\end{minipage}

\vspace{0.8em}

\begin{minipage}{\textwidth}
\centering
\textbf{(b)}\par\vspace{0.25em}
\includegraphics[width=\linewidth]{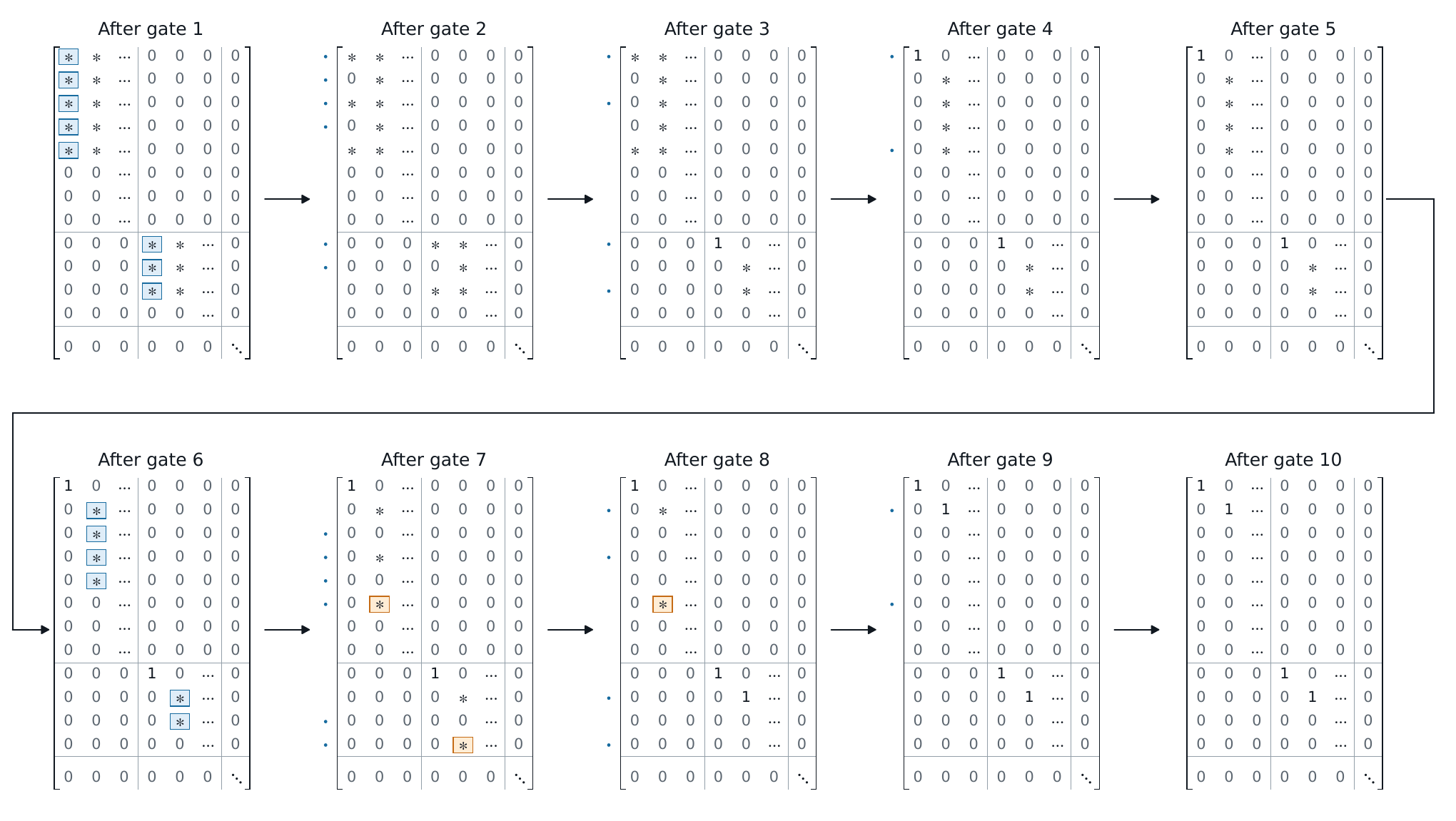}
\end{minipage}
\caption{\textbf{Block-isometry synthesis.}
(a) Column elimination using diagonal phase layers $\Delta$ and controlled $R_y$ rotations; reversing the circuit synthesizes the isometry.
In this notation, blank white boxes represent multiplexed rotations, while white boxes marked with bit labels mean the rotation is applied only when the qubit matches the bit value shown in the box.
(b) Ideal elimination of the first two columns of a $5\times3$ block and a $3\times2$ block, padded to eight and four rows, respectively.
The upper and lower sequences clear columns zero and one.
Blue boxes mark entries whose phases are cancelled, blue dots mark rows participating in a rotation, and orange boxes mark nonzero entries in padding.
Stars denote nonzero entries; ellipses abbreviate the remaining columns of each full unitary block and the trailing identity block.
Gates 5 and 10 are identity levels.
The larger block's third column is cleared by a subsequent sequence, not shown.}
\label{fig:block-isometry}
\end{figure}

The orange entries in Fig.~\ref{fig:block-isometry}(b) illustrate why the elimination may require access to padding: clearing one column can move a different column's amplitude into padded rows.
The first column's tree preserves the original row support of every supplied column, but later columns can populate padding in a structured way.
Writing $q_i=\lceil\log_2M_i\rceil$, it suffices to track support rounded up to groups of $2^{q_i}$ rows,
\begin{equation}
  N_i^{\mathrm{eff}}
  =2^{q_i}\left\lceil\frac{N_i}{2^{q_i}}\right\rceil
  \leq R_i.
  \label{eq:block_isometry_support}
\end{equation}
The lowest $q_i$ address bits mix rows only within these groups.
At higher levels, every destination $c<M_i$ has target-bit value zero, so if a pair straddles the boundary of this support then its ``outside'' member already has amplitude in the correct member and the rotation is skipped.

Subset-restricted QROAM supplies one angle table for each nontrivial phase layer or rotation level, combining all active blocks in a single lookup.
Let $\mathcal P_c$ be the set of full row addresses requiring a nonzero phase at column $c$, and let $\mathcal S_{c,t}$ be the set of pair addresses requiring a nonzero rotation at level $t$.
Phase lookups use all $n$ address bits and are confined to uncleared rows below $N_i^{\mathrm{eff}}$ in active blocks.
Rotation lookups use the upper $n-t-1$ bits: deleting the target bit and the fixed lower bits yields a unique address per selected pair, and the lookup is enabled only when the lower $t$ bits match $c$.
Completed blocks and zero-angle records are omitted, leaving gaps that subset-restricted QROAM handles directly.
We compute these tables from the ideal elimination sequence, round each angle once to $b$ bits, and retain the same lookup subsets after rounding.

After each loaded angle is applied, we measure all QROAM data registers in the $X$ basis and reuse them.
We can follow similar strategies of Ref.~\cite{berry2024rapid} which delays the uncomputation towards the last layer, leaving only the forward computation cost for each layer.
To achieve this, the measurements introduce known signs on computational-basis amplitudes.
Rather than correcting after every lookup, we track these signs classically.
Before a subsequent real rotation, we flip the sign of its angle if the two basis states in the selected pair have different tracked signs; if their signs agree, the angle is left unchanged.
This passes the accumulated signs through the rotation without changing the intended operation (diagonal phase layers already commute with them).
Since each rotation record addresses one selected pair, this adjustment changes only the sign of the angle and does not change the lookup subset.

A final diagonal correction removes the accumulated signs on the entire register, preserving the chosen unitary completion in either circuit direction (this is required to ensure the entire block-encoding unitary is self-inverse).
Let $\mathcal F$ be a sufficient set of addresses for this correction.
It must include signs induced by measuring unused QROAM words, which can occur even when a selected angle is zero.
For each lookup with computation parameter $\lambda$, we therefore expand its retained subset to whole aligned groups of $\lambda$ lookup addresses (restoring the deleted bits for rotation lookups by inserting the lower $t$ bits of $c$ and both values of the target bit).
Taking the union over all lookups defines $\mathcal F$, including any affected padding, completed blocks, or unused addresses.
The final correction table records the accumulated sign on each address in $\mathcal F$ and applies no correction elsewhere.

To count Toffolis, let $\ell$ index the nonempty phase and rotation lookups, with subset $\mathcal S_\ell$ and address width $n_\ell$ (equal to $n$ for phases and $n-t-1$ for a level-$t$ rotation).
Each lookup has an independently chosen power-of-two computation parameter $1\leq\lambda_\ell\leq2^{n_\ell}$.
Lower-bit conditions are built incrementally as shared temporary ANDs~\cite{gidney2018halving}, starting from the external control.
If $d_c$ is the highest level $t$ with a retained rotation at column $c$ (or zero if there are none), these controls require $d_c$ Toffolis to compute and only measurements and Clifford corrections to erase.
Writing $M_*=\max_i M_i$, the scheduled cost in either circuit direction is
\begin{equation}
  \begin{aligned}
  C_{\mathrm{isometry}}
  ={}&\sum_\ell\left[
  C_{\operatorname{QROAM}}
  (\mathcal S_\ell,n_\ell,b,\lambda_\ell)+(b-2)
  \right]\\
  &+\sum_{c=0}^{M_*-1}d_c
  +C_{\operatorname{QROAM}^{-1}}(\mathcal F,n,\lambda').
  \end{aligned}
  \label{eq:block_isometry_cost}
\end{equation}
The three contributions are the angle lookups and applications, the shared controls, and the final sign correction.
The last term uses the measurement-based QROAM uncomputation primitive with an independent power of two $2\leq\lambda'\leq2^n$, and it vanishes when $\mathcal F$ is empty.
No ordering condition between $\lambda'$ and the computation parameters is needed because $\mathcal F$ already covers the full support of the deferred signs.
Ancillas are reused throughout; the cost excludes preparation of the phase-gradient state and classical table processing.

For $B$ blocks with $N$ padded rows and $M$ columns each, every column needs one phase lookup over $NB$ addresses and, at successive tree levels, rotation lookups over $NB/2,NB/4,\ldots$ pair addresses.
In the T-optimal variant, each lookup uses its own optimal $\lambda_\ell$, so a column costs approximately $2\sqrt{bNB}\,\bigl(1+2^{-1/2}+2^{-1}+\cdots\bigr)=2(2+\sqrt2)\sqrt{bNB}\simeq 7\sqrt{bNB}$ Toffolis.
The $M$ columns cost $7\sqrt b\,M\sqrt{NB}$ Toffolis, with $\sqrt{bNB}$ workspace qubits set by the phase lookup.
In the Q-optimal variant, every lookup uses $\lambda_\ell=1$: a column costs $NB$ Toffolis for its phases and $NB(\tfrac12+\tfrac14+\cdots)\simeq NB$ for its rotations, so the $M$ columns cost $2MNB$ Toffolis with $3b+\log_2(NB)$ qubits.
The final sign correction, with $\lambda'$ optimized as in Appendix~\ref{app:qroam}, adds approximately $NB/\lambda'$ Toffolis.
These are the block-isometry entries of Table~\ref{tab:primitives}(c).

\subsection{Block-state preparation}

In the synthesis of exchange proto-term, we will use the block-state preparation, which is the single-column case of block-isometry synthesis: for each block $i$, an operator $S_i$ satisfies $S_i\ket{0}=\ket{\psi_i}\in\mathbb C^{N_i}$ and can be implemented with the same binary-tree pattern of rotations, but without the lower-bit controls needed to select among multiple columns.

When the state dimension is much smaller than the number of blocks, it is more efficient to ``load all angles at once'' (rather than querying a new angle at every tree node).
We therefore use the load-all construction of Sec.~III.C (Fig.~10) of Ref.~\cite{clader2022quantum}.
Concretely, we pad all states to a common power-of-two dimension $R=2^p\geq\max_i N_i$ so that a single fixed rotation tree can be reused for every block.
One QROAM query addressed by the block index then loads the full set of angles needed for that tree: the $R-1$ magnitude angles and the $R$ phases, as a single word of width $W=(2R-1)b$.
Controlled swaps route the appropriate angle to the active rotation at each tree level (and route the final phase); after applying the rotations, we reverse the swaps and erase the loaded word using measurement-based QROAM uncomputation.

Writing $\mathcal B=\{0,\ldots,B-1\}$ and $n_B=\lceil\log_2B\rceil$, the Toffoli cost is
\begin{equation}
  \begin{aligned}
  C_{\mathrm{state}}
  ={}&C_{\operatorname{QROAM}}(\mathcal B,n_B,W,\lambda)\\
  &+C_{\operatorname{QROAM}^{-1}}(\mathcal B,n_B,\lambda')
  +2b(2R-p-2)+(p+1)(b-2).
  \end{aligned}
  \label{eq:block_state_preparation_cost}
\end{equation}
Here $\lambda'\geq\lambda$ ensures that the final uncomputation accounts for signs induced by measurements on unused QROAM words.
The explicit $2b(2R-p-2)\simeq4bR$ term comes from routing the angle word to the rotation sites and then restoring it; the rotations themselves and any shared controls contribute only $(p+1)(b-2)$, for the $p$ magnitude levels and the final phase.
The inverse applies the same rounded rotations in reverse order and has the same cost.

For $B$ states of dimension $N=R$, the T-optimal variant costs $(2\sqrt{WB}-W)+2\sqrt B+4bN\simeq2\sqrt2\sqrt{bNB}+2\sqrt B+2bN$ Toffolis, with $W=(2N-1)b\simeq2bN$, which rounds to the $3\sqrt{bNB}+2bN$ of Table~\ref{tab:primitives}(c); its peak ancilla count is $\max\{W\lambda,\lambda'\}+O(b+n_B+p)$, with $W\lambda\simeq\sqrt{2bNB}$.
Since the loaded word alone requires $W\simeq2bN$ ancilla qubits, even at $\lambda=1$, the Q-optimal variant instead uses the single-column case of the block-isometry synthesis of Appendix~\ref{app:block-isometry}: a phase lookup over $NB$ addresses and rotation-angle lookups over $NB/2,NB/4,\ldots$ addresses, each with $\lambda=1$ and one $b$-bit angle register.
It costs $2NB$ Toffolis and $3b+\log_2(NB)$ ancilla qubits, with the measurement-based uncomputation using the optimized $\lambda'$ of Appendix~\ref{app:qroam}.
These are the block-state-preparation entries of Table~\ref{tab:primitives}(c).

\section{Synthesis of the space-group transformation}
\label{app:S}

This appendix describes the circuit for the THC-side basis change $S^{\mathrm{THC}}$ of Eqs.~\eqref{eq:be_zeta_sg_main}--\eqref{eq:be_chi_sg_main}.
Its columns form a space-group-adapted basis for the representation $D_g^{\mathrm{THC}}$ on the THC index $(\kk,\mu)$.
A space-group irrep $\lambda=(f,\alpha)$ is labelled by a momentum star $f$, with representative momentum $\kk_f$, and an irrep $\alpha$ of the little group of $\kk_f$~\cite{bradley2010mathematical}; its dimension $m_\lambda$ is the number of star members times the dimension of $\alpha$.

The key idea is to avoid synthesizing an independent unitary at every momentum.
Since $D_g^{\mathrm{THC}}$ maps the THC space at $\kk_f$ to that at $g\kk_f$, the basis vectors at every star member follow from those at the representative by applying the symmetry itself, i.e.\ by a permutation of interpolation points and a Bloch phase.
We therefore (i) synthesize unitary blocks only at the star representatives, using the block-unitary primitive of Appendix~\ref{app:eigen}, and (ii) use QROAM to convert coherently between the space-group labels and the momentum labels needed to apply the appropriate permutation and phase.
The circuit has five stages, described below.

\paragraph{Registers and target map.}
Let $B$ be the number of irrep blocks and $N_f$ the number of momentum stars.
The input register $\ket L$ encodes $(\lambda,c)$, where $c<n_\lambda$ is the copy index.
Block $\lambda$ occupies $\bar n_\lambda$ addresses, with $\bar n_\lambda$ the smallest power of two at least $n_\lambda$ for $\boldsymbol{\chi}^X$ and the smallest even integer at least $n_\lambda$ for $\boldsymbol{\zeta}^X$.
The partner register $\ket a=\ket\kappa\ket\beta$ indexes the $m_\lambda$ copies in $I_{m_\lambda}$: it contains the star-member index $\kappa$ and the row index $\beta$ of the little-group irrep $\alpha$.
The target map is
\begin{equation}
  \ket{L}\ket{a}\ \longrightarrow\
  \ket{\kk}\sum_{\mu}S^{\mathrm{THC}}_{(\kk,\mu),(\lambda,a,c)}\ket{\mu}.
  \label{eq:S_goal}
\end{equation}

\paragraph{Stage 1: space-group labels.}
We first recover the space-group labels from $L$.
With offsets $s_\lambda=\sum_{j<\lambda}\bar n_j$ and compact offsets $t_\lambda=\sum_{j<\lambda}n_j$, the staircase shift $L\mapsto L+\sum_{j:L\geq t_{j+1}}(\bar n_j-n_j)$ separates the block and copy indices.
Ordering the power-of-two blocks by decreasing size aligns each $s_\lambda$ to a multiple of $\bar n_\lambda$, so the lowest $\log_2\bar n_\lambda$ bits already encode $c$ for $\boldsymbol{\chi}^X$.
A coherent binary search over the $B-1$ block boundaries recovers $\lambda$ in $\lceil\log_2B\rceil$ levels, using comparators of width $n=\lceil\log_2\sum_\lambda\bar n_\lambda\rceil$.
The first threshold is constant and the second is selected by CNOTs; only levels $r\geq3$ require lookups, with $2^{r-1}$ entries per level and $O(B)$ entries in total.
Two $B$-entry lookups map $\lambda$ to $(f,\alpha)$ and erase $\lambda$, while an offset lookup and subtraction recover $c=L-\sum_{j<\lambda}n_j$.
Schur's lemma makes these lookups independent of $a$, so padding $a$ into fixed-width $\kappa$ and $\beta$ subregisters requires no gates.
The label-conversion cost is $O(n\log B+\sqrt{B\log B})$ Toffolis.

\paragraph{Stage 2: momentum.}
The momentum is reconstructed as $\kk=A_g\kk_f$, where $\kk_f$ is loaded from an $N_f$-entry table, $g$ is the coset representative selected by $\kappa$, and $A_g$ is the action of $g$ on momenta ($g\kk\equiv A_g\kk$).
In cubic coordinates, $A_g$ is a signed permutation of the three components, giving a specialized cost of $O(|P|+\sqrt{N_f\log N_k})$, where $|P|$ is the point-group order.
The tabulated implementation instead uses a $(g,f)$-indexed lookup in primitive reciprocal coordinates.

\paragraph{Stage 3: representative blocks.}
At each star representative, the little group permutes the interpolation points, which therefore separate into little-group orbits.
In this orbit basis, the representative block of $S^{\mathrm{THC}}$ is
\begin{equation}
  S^{\mathrm{THC}}_{f}=\bigoplus_\omega U_\omega,
  \qquad U_\omega\in\mathbb C^{N_\omega\times N_\omega},
  \label{eq:S_star_blocks}
\end{equation}
where the orbit sizes $N_\omega$ divide $|P|$ and sum to $\NTHC$ within each star.
We apply the block-unitary synthesis of Eq.~\eqref{eq:block-unitary-cost}, with lookups addressed by $(f,\omega)$.
The circuit has $\max_\omega N_\omega\leq|P|$ layers, and layer $\ell$ loads $\sum_{N_\omega>\ell}\lfloor N_\omega/2\rfloor$ rotation pairs.
Subset lookups leave completed orbits and unused addresses unchanged.
The synthesis cost is $O(|P|\sqrt{N_f\NTHC b})$; its mesh dependence enters only through the number of stars $N_f$.

\paragraph{Stage 4: remaining star members.}
The block at $\kk=A_g\kk_f$ is obtained by applying the corresponding block of $D_g^{\mathrm{THC}}$ to the representative block (induction):
\begin{equation}
  S^{\mathrm{THC}}_{\kk}=\Pi(g)\,\Lambda(g,\kk)\,S^{\mathrm{THC}}_{f},
  \qquad
  \Lambda_{\mu\mu}(g,\kk)=e^{-\ii\kk\cdot\LL_\mu(g)}.
  \label{eq:S_induced_blocks}
\end{equation}
Here $\Pi(g)$ maps $\mu$ to $\mathrm{perm}[g,\mu]$, and the lattice vector $\LL_\mu(g)$ satisfies $g\,\rr_\mu=\rr_{\mathrm{perm}[g,\mu]}+\LL_\mu(g)$.
Two geometry tables of size $|P|\times\NTHC$ store these permutations and lattice vectors, independently of $\kk$.
The Bloch phase factorizes over the three axes, $e^{-\ii\kk\cdot\LL_\mu(g)}=\prod_{j}e^{-2\pi\ii k_jL_{\mu,j}(g)/N_j}$ for a $N_1\times N_2\times N_3$ mesh.
For each axis, a lookup over the $N_j$ mesh values loads the $b$-bit phase $k_j/N_j$, which is then added into the phase-gradient register once per bit of $L_{\mu,j}(g)$, controlled on that bit (subtracted for the sign bit); the weights $2^i$ are bit shifts.
It cancels across the diagonal central factor of the direct term but remains in the exchange term.
The induction cost is $O(\sqrt{|P|\NTHC\log\NTHC}+b\sum_jw_j)$, where $w_j$ is the bit width of $L_{\mu,j}(g)$, fixed by the geometry.

\paragraph{Stage 5: uncomputation.}
Finally, $\kk$ uniquely determines $(f,\kappa)$, so these temporary labels can be erased by measurement-based QROAM uncomputation over $N_k$ addresses, at a cost of $O(\sqrt{N_k})$ Toffolis.
Alternatively, the canonical representative $\kk^*=\min_gA_g\kk$ can be computed over the $|P|$ group operations at cost $O(|P|\log N_k)$; the lookup is cheaper for the meshes considered here.

\paragraph{Cost.}
Table~\ref{tab:diamond-symmetry-synthesis} reports the cost of one application of $S^{\mathrm{THC}}$.
It increases by a factor of $1.24$ between $2\mathbin{\times}2\mathbin{\times}2$ and $6\mathbin{\times}6\mathbin{\times}6$, while $N_k$ increases by a factor of $27$.
The representative unitaries have dimension at most $|P|$, but their number grows with the number of stars $N_f$, which accounts for most of the mesh dependence; the label conversion, the momentum lookup, and the final momentum-addressed uncomputation contribute the rest.

\begin{table}[H]
\centering\small
\begin{ruledtabular}
\begin{tabular}{rrrr}
  $N_k$ & $N_f$ & $B$ & Toffolis for $S^{\mathrm{THC}}$\\
  \hline
  8 & 3 & 20 & 10936\\
  64 & 8 & 36 & 12492\\
  216 & 16 & 57 & 13548\\
  512 & 29 & 87 & 15131\\
  1000 & 47 & 124 & 16545\\
\end{tabular}
\end{ruledtabular}
\caption{\textbf{Illustrative space-group synthesis costs for diamond.}
Counts for one application of $S^{\mathrm{THC}}$ from the implemented five-stage model at $b=32$ and $\NTHC=208$ use the selected-point geometry to determine the momentum stars, irreps, and little-group orbits, with even-padded copy blocks.
Momentum reconstruction uses a $(g,f)$ lookup in primitive reciprocal coordinates.
The counts exclude phase-gradient-state preparation.}
\label{tab:diamond-symmetry-synthesis}
\end{table}

\section{Operator norm of the BSE Hamiltonian}
\label{app:bse-norm}

The goal of this appendix is to explain why the physical BSE Hamiltonian has an $O(1)$ operator norm in the thermodynamic limit, at fixed excitation number and basis resolution.
We consider a gapped system and assume that the occupied and retained virtual spaces admit exponentially localized orbitals, with localization lengths independent of system size~\cite{brouder2007exponential,marzari2012maximally}.

Assuming bounded eigenvalues of the Fock blocks, we only need to show that the individual direct and exchange Hamiltonians have $O(1)$ norm.
The full $m$-exciton Hamiltonian then also has $O(1)$ norm in system size, up to a $\operatorname{poly}(m)$ factor from the number of Fock and two-body terms.

For the direct interaction, exponential localization gives
\begin{equation}
  (pq|W|rs)
  = O\!\left(e^{-a d(p,q)}e^{-a d(r,s)}\right),
  \label{eq:direct-locality}
\end{equation}
where $d(p,q)$ is the separation between the orbital centers.
Here we assume that short-range interaction matrix elements remain bounded at fixed basis resolution.
Viewed as a matrix with row index $(p,r)$ and column index $(q,s)$, the exponential decay means that each row couples appreciably to only $O(1)$ entries.
The direct Hamiltonian therefore has $O(1)$ operator norm, even though the interaction between the two orbital pairs remains long-ranged.

For exchange, the occupied--virtual transition densities are charge-neutral because the occupied and virtual orbitals are orthogonal.
Their leading long-range interaction is therefore dipole--dipole, giving
\begin{equation}
  (pq|V|rs)
  = O\!\left(
  \frac{e^{-a d(p,q)}e^{-a d(r,s)}}{d(pq,rs)^3}
  \right)
  \label{eq:exchange-locality}
\end{equation}
at large separation $d(pq,rs)$ between the two orbital pairs.
Including its angular dependence, the dipole--dipole interaction transforms into a momentum-space kernel~\cite{gatti2013exciton} proportional to
\begin{equation}
  \frac{\boldsymbol{Q}\boldsymbol{Q}}
  {|\boldsymbol{Q}|^2}-\frac{I}{3},
  \label{eq:dipole-kernel}
\end{equation}
which is bounded independently of system size.
Together with the bounded short-range contribution, this gives an $O(1)$ norm for the exchange Hamiltonian viewed as a matrix between transition-pair indices.
Thus the full BSE Hamiltonian has $O(1)$ operator norm for fixed $m$.

\section{THC rank and subnormalization}
\label{app:thc-scaling}

The goal of this appendix is to explain why the THC-based block-encoding can retain the $O(1)$ physical norm established in Appendix~\ref{app:bse-norm} while using an $O(N)$ THC rank.
We use the same assumptions of a gapped system, fixed basis resolution, and exponentially localized occupied and virtual orbitals.
To guide the argument, we exhibit an explicit, although unoptimized, construction at fixed approximation accuracy.
This provides a baseline that THC optimization can further improve by reducing the rank and subnormalization.

For either interaction $X\in\{W,V\}$ and any ERI block, consider the exact THC representation in which the auxiliary index is the orbital-pair label $(PQ)$:
\begin{equation}
  \begin{split}
  (pq|X|rs) = \sum_{(PQ),(RS)}
  &\chi^{X,A}_{p,(PQ)}\chi^{X,B}_{q,(PQ)}\\
  &{}\times\chi^{X,C}_{r,(RS)}\chi^{X,D}_{s,(RS)}
  \zeta^X_{(PQ),(RS)},
  \end{split}
  \label{eq:pair-index-thc}
\end{equation}
with
\begin{align*}
  \chi^{X,A}_{p,(PQ)} &= \delta_{pP},\\
  \chi^{X,B}_{q,(PQ)} &= \delta_{qQ},\\
  \zeta^X_{(PQ),(RS)} &= (PQ|X|RS),
\end{align*}
and similarly for $\chi^{X,C},\chi^{X,D}$.
Here $A,B,C,D$ label the four tensor-leg selector maps.

This representation is exact but has an unhelpful rank.
Exponential locality allows us to discard orbital pairs whose members are far apart~\cite{prodan2005nearsightedness}, with an error that decreases exponentially with the retained separation.
At fixed accuracy, each $P$ connects to only $O(1)$ values of $Q$, and vice versa.
Hence the number of surviving pair indices scales as $O(N)$.

The same structure keeps the operator norms of the $\chi^{X,A},\chi^{X,B},\chi^{X,C},\chi^{X,D}$ matrices $O(1)$: each row contains only $O(1)$ nonzero entries, and each column selects a single orbital.
For the direct channels, $\zeta^W$ enters the circuit as a diagonal operator whose norm is its largest absolute entry, which remains $O(1)$.
For exchange, $\zeta^V$ is the physical exchange matrix restricted to the retained transition pairs, so its operator norm remains $O(1)$ by Appendix~\ref{app:bse-norm}.
The resulting products of factor norms therefore give $O(1)$ primitive subnormalizations.

This explicit construction demonstrates that $O(N)$ THC rank is sufficient to achieve $O(1)$ subnormalization at fixed excitation number and accuracy.
The construction is not intended to give optimal constants; further THC optimization can reduce the required rank and subnormalization while maintaining the target accuracy.

\section{Calibration of THC fitting accuracy}
\label{app:fit_THC}

The bare and screened interactions require a common criterion for selecting THC decompositions with sufficient accuracy.
For the bare interaction $V$, the MP2 correlation-energy error provides an energy-based measure of the fitting error in the occupied--virtual block $(ov|vo)$ \footnote{Strictly speaking MP2 only uses $(ov|ov)$ instead of $(ov|vo)$, however for systems satisfying time-reversal symmetry, they are equivalent up to an index permutation and complex conjugation, therefore we use this principle to convert $(ov|vo)$ to $(ov|ov)$ for the MP2 energy evaluation.}, as used in Ref.~\cite{rubin2023faulttolerant}.
The screened interaction $W$ entering the BSE does not have a corresponding standard MP2 energy-error metric.
We therefore calibrate a relative-integral-error tolerance using the bare interaction and transfer that tolerance to the screened interaction.
In the following we state the principle using diamond as an example, while other materials just follow the same principle.

\begin{figure}[htbp]
\centering
\includegraphics[width=0.68\textwidth]{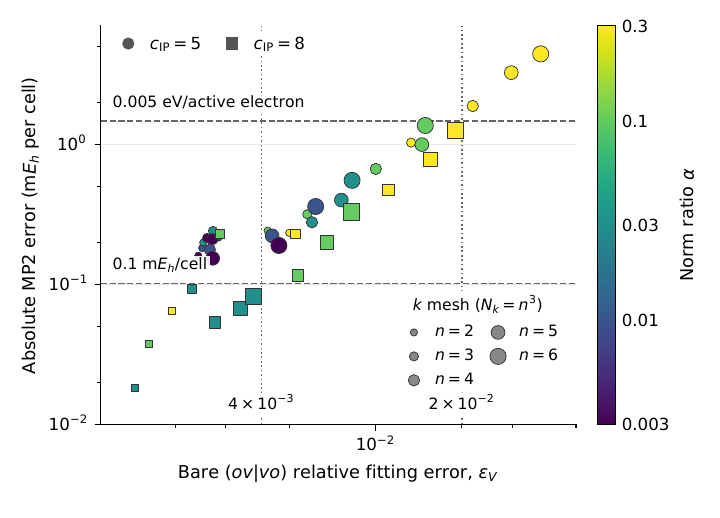}
\caption{Calibration of THC fitting accuracy in diamond.
Absolute MP2 energy error per primitive cell versus the relative THC fitting error $\epsilon_V$ of the bare $(ov|vo)$ block.
Marker size indicates the mesh edge $n$, with $N_k=n^3$, and color indicates the norm-ratio parameter $\alpha$ used in the THC optimization, on a logarithmic color scale.
Circles and squares indicate $c_{\mathrm{IP}}=5$ and $c_{\mathrm{IP}}=8$, respectively.
All sampled optimization settings are included.
Horizontal dashed lines mark $0.1\,\mathrm{m}E_h$ per cell and $0.005\,\mathrm{eV}$ per active electron, the latter equivalent to $1.47\,\mathrm{m}E_h$ per diamond cell when counting its $8$ active electrons.
Vertical dotted lines mark the relative-error targets $4\times10^{-3}$ and $2\times10^{-2}$ used in the calculations for the $(ov|W|vo)$ projection of the full screened THC fit.}
\label{fig:thc-mp2-calibration}
\end{figure}

Figure~\ref{fig:thc-mp2-calibration} shows the diamond calibration data for $n\times n\times n$ meshes with $n=2,\ldots,6$ and THC rank as $N_\text{THC}=c_{\mathrm{IP}} N_\text{AO}$, $c_{\mathrm{IP}}=5$ and $8$, including varying $\alpha$ from 0.003 to 0.3 that balances the fitting accuracy and subnormalization via the optimization loss function (see Eq.~\ref{eq:loss}).
The MP2 energy error per primitive cell generally increases with the relative error of the bare $(ov|vo)$ block, although the relation has scatter and depends on the mesh and fitting parameters.
We use this relation to estimate the relative-error scale associated with each of the two energy-accuracy criteria used in the main text.
For comparison with the full-Hamiltonian quantum simulation references, an MP2 error of $0.1\,\mathrm{m}E_h$ per cell corresponds to relative integral errors of $\sim 10^{-3}$; we use a target of $4\times10^{-3}$.
For the realistic BSE estimates, the target is $0.005\,\mathrm{eV}$ per active electron.
Diamond has $8$ active electrons in its two-carbon primitive cell, excluding the core electrons represented by the pseudopotential~\cite{goedecker1996separable,hartwigsen1998relativistic}.
This criterion is therefore $0.040\,\mathrm{eV}\simeq1.47\,\mathrm{m}E_h$ per cell.
The corresponding relative-error scale is a few percent; we use a target of $2\times10^{-2}$.
For zinc-blende CdSe and monolayer MoSe$_2$, the target for the direct term is the relative error that the bare exchange fit reaches at the MP2 criterion of each material (median over meshes).
MoSe$_2$'s direct term is fitted to the bare interaction.

For $X=V$ or $W$, write the relative error in the projected block as
\begin{equation}
  \epsilon_X=
  \frac{\left\|X^{\mathrm{THC}}_{ov,vo}-X_{ov,vo}\right\|_2}
  {\left\|X_{ov,vo}\right\|_2},
  \label{eq:thc-projected-error}
\end{equation}
using the same relative-error norm as in the main text.
We fit the full screened ERI tensor and evaluate $\epsilon_W$ on its $(ov|W|vo)$ projection.
The two screened-interaction fitting targets are thus $\epsilon_W\lesssim4\times10^{-3}$ and $\epsilon_W\lesssim2\times10^{-2}$, respectively.
This projection is used to assess the accuracy of the full screened fit, rather than restricting the fit to that block.
The transfer from the bare-integral calibration is an empirical fitting criterion; it does not introduce an MP2 energy for the screened interaction.

\section{Extrapolation details}
\label{app:baselines}

This appendix describes how the first- and second-quantized full-Hamiltonian estimates in Fig.~\ref{fig:optimized-literature-comparison} are extended beyond the meshes reported in Refs.~\cite{rubin2023faulttolerant,berry2024quantum}.
At every mesh we evaluate the published circuit formulas, and fit only the quantities that the sources do not give in closed form.

\paragraph{Precision.}
All quantum methods in Fig.~\ref{fig:optimized-literature-comparison} use $\epsilon_{\mathrm{QPE}}=0.1\,\mathrm{eV}\simeq3.675\,\mathrm{m}E_h$ for one eigenvalue estimate: an excitation energy for BSE or a total supercell energy for the full Hamiltonian.
Constant overheads for combining two total-energy estimates into an excitation energy are omitted.
The published reference estimates use $\epsilon_{\mathrm{ref}}=1.6\,\mathrm{m}E_h$; changing the numerical tolerance therefore rescales the number of walk steps by $\epsilon_{\mathrm{ref}}/\epsilon_{\mathrm{QPE}}\approx0.435$.
For second quantization~\cite{rubin2023faulttolerant}, the additional conversion from per-primitive-cell to total-supercell accuracy contributes a factor of $N_k$, giving an overall walk-count factor of approximately $0.435N_k$ relative to the published convention.
No such $N_k$ conversion is needed for first quantization~\cite{berry2024quantum}.
Circuit costs and phase-estimation register sizes are evaluated at the new precision.

\paragraph{Second quantization.}
We use the symmetry-adapted $k$-point THC construction of Ref.~\cite{rubin2023faulttolerant} with $52$ spin orbitals per primitive cell and $\NTHC=8\NAO=208$.
The subnormalization $\lambda_{\mathrm{THC}}$ depends on the optimized THC factors and is reported only for diamond meshes up to $2\times2\times2$.
The THC sum contains $O(N_k^3)$ momentum combinations, but each effective coefficient is expected to decay as $O(N_k^{-1})$, giving
\begin{align}
  \lambda_{\mathrm{THC}}=O(N_k^2)
  \label{eq:sq_lambda_scaling}
\end{align}
at fixed $\NTHC$~\cite{rubin2023faulttolerant}.
The fits in Fig.~9(b) of Ref.~\cite{rubin2023faulttolerant} are consistent with this: over the benchmark set, $\lambda_{\mathrm{THC}}$ grows slightly faster than quadratically for symmetry-adapted THC and slightly slower for supercell THC.
We therefore take $\lambda_{\mathrm{THC}}$ to be quadratic, so that $\lambda_{\mathrm{THC}}/N_k$ is linear in $N_k$, and fix the line with the two cubic meshes $1\times1\times1$ and $2\times2\times2$:
\begin{align}
  \frac{\lambda_{\mathrm{THC}}}{N_k}
  =557.5808N_k-214.5719.
  \label{eq:sq_lambda_fit}
\end{align}
The walk cost and logical-qubit count then follow from the published Appendix-D circuit formulas at each mesh; the qubit count includes the phase-estimation register required by the larger number of walk steps.
At $10\times10\times10$, this gives $\lambda_{\mathrm{THC}}=5.6\times10^8$, $1.6\times10^6$ Toffolis per walk, $4.1\times10^5$ logical qubits, and $3.7\times10^{17}$ Toffolis.
At fixed THC rank ratio and precision, the total Toffoli count scales as
\begin{align}
  \widetilde O(\lambda_{\mathrm{THC}}N_k\NAO).
  \label{eq:sq_scaling}
\end{align}

\paragraph{First quantization.}
We use the pseudopotential plane-wave construction of Ref.~\cite{berry2024quantum}, anchored at its diamond $3\times3\times3$ instance: $\lambda=8.402\times10^6$, $23{,}576$ Toffolis per walk, and $1.94\times10^{14}$ Toffolis in total at $1.6\,\mathrm{m}E_h$.
We hold the $80\,\mathrm{Ry}$ plane-wave resolution of that instance fixed, so that the extrapolation isolates the mesh dependence.
For a $k\times k\times k$ crystal, the momentum register then has width
\begin{align}
  n(k)=\left\lceil\log_2(21k+1)\right\rceil.
  \label{eq:fq_grid}
\end{align}
At each $k$, we evaluate the kinetic, electron--electron, local-pseudopotential, and nonlocal-pseudopotential subnormalizations and the block-encoding cost from the source formulas on this grid; no fit is applied.
The logical-qubit count adds the system register, the phase-estimation register, and $8000$ arithmetic ancillas, the last being an extrapolation assumption.
At $10\times10\times10$, this gives $\lambda=1.2\times10^{10}$, $8.1\times10^5$ Toffolis per walk, $2.0\times10^5$ logical qubits, and $4.2\times10^{18}$ Toffolis.
At fixed composition, pseudopotential parameters, and precision, the total Toffoli count scales as
\begin{align}
  \widetilde O\!\left(
  N_k^2n_{\mathrm{PW}}^{2/3}
  +N_k^3n_{\mathrm{PW}}^{1/3}+N_k^3\right),
  \label{eq:fq_scaling}
\end{align}
where $n_{\mathrm{PW}}$ is the number of plane waves per primitive cell.
The three terms arise from the kinetic, Coulomb, and pseudopotential subnormalizations, respectively, and give cubic scaling in $N_k$ at fixed $n_{\mathrm{PW}}$.

\clearpage

\section{Resource data and tables}
\label{app:resource-tables}

Table~\ref{tab:material-resources} extends Fig.~\ref{fig:optimized-exciton-comparison} to the smaller $k$-point meshes, and lists the data used to extrapolate to $10\times10\times10$.
The settings are those of the figure: Toffoli-optimized walks, $0.1\,\mathrm{eV}$ QPE precision, and THC fits at $0.005\,\mathrm{eV}$ MP2 error per active electron, with rank $c_{\mathrm{IP}}=5$ for the exchange terms and $c_{\mathrm{IP}}=4$ for the direct terms.
For zinc-blende CdSe, the fits use $c_{\mathrm{IP}}=4$ for the exchange term and $c_{\mathrm{IP}}=5$ for the direct term, and for monolayer MoSe$_2$ they use $c_{\mathrm{IP}}=4$ for both terms, with the direct term fitted to the bare Coulomb interaction.
Table~\ref{tab:material-resources-2d} gives the MoSe$_2$ resources on $k\times k$ meshes up to the $32\times32$ mesh of Fig.~\ref{fig:optimized-exciton-comparison}.
Rows marked (SG) use space-group adaptation and the same ranks as the rows without it.
For diamond, MgO, BN, and AlN, the exchange terms use the symmetric selection of interpolating points at $c_{\mathrm{IP}}=5$ and the direct terms use $c_{\mathrm{IP}}=4$.
For CdSe and MoSe$_2$, the symmetric selections have the ranks of the fits without space-group adaptation, with $154$ (exchange) and $201$ (direct) points for CdSe and $214$ and $215$ points for MoSe$_2$.
The direct and exchange terms have different irrep blocks and separate $S^{\mathrm{THC}}$, giving ten applications of $S^{\mathrm{THC}}$ per walk (four from direct and six from exchange).
For diamond, $\lambda$ (SG) is the subnormalization of the space-group-adapted fits.
The diamond (SG) Toffoli and qubit counts use this fitted subnormalization.
For MgO, BN, AlN, CdSe, and MoSe$_2$, the (SG) rows combine the space-group-adapted walk with the subnormalization from the fit without space-group adaptation.

\begin{table*}[t]
\centering\small
\setlength{\tabcolsep}{4pt}
\begin{ruledtabular}
\begin{tabular}{llrrrrrr}
  \textbf{System} & & $3\times3\times3$ & $4\times4\times4$ & $5\times5\times5$ & $6\times6\times6$ & $8\times8\times8$ & $10\times10\times10$\\
  \hline
  Diamond, $m=2$ & $\lambda$ & $46.2$ & $51.2$ & $55.6$ & $60.8$ & $68.3^{\ast}$ & $75.1^{\ast}$\\
  & $\lambda$ (SG) & $47.4$ & $50.0$ & $52.2$ & $54.9$ & $58.4^{\ast}$ & $61.5^{\ast}$\\
  & Toffolis per walk & $5.9{\times}10^{5}$ & $8.8{\times}10^{5}$ & $1.3{\times}10^{6}$ & $1.6{\times}10^{6}$ & $2.5{\times}10^{6}$ & $3.5{\times}10^{6}$\\
  & Toffolis per walk (SG) & $1.6{\times}10^{5}$ & $2.8{\times}10^{5}$ & $3.2{\times}10^{5}$ & $4.0{\times}10^{5}$ & $6.7{\times}10^{5}$ & $8.6{\times}10^{5}$\\
  & Logical qubits & $4{,}260$ & $4{,}265$ & $8{,}365$ & $8{,}370$ & $16{,}566$ & $16{,}571$\\
  & Logical qubits (SG) & $4{,}260$ & $4{,}265$ & $8{,}365$ & $8{,}370$ & $16{,}566$ & $16{,}571$\\
  & Total Toffolis & $1.2{\times}10^{10}$ & $1.9{\times}10^{10}$ & $3.0{\times}10^{10}$ & $4.2{\times}10^{10}$ & ${7.4{\times}10^{10}}^{\ast}$ & ${1.1{\times}10^{11}}^{\ast}$\\
  & Total Toffolis (SG) & $3.3{\times}10^{9}$ & $5.9{\times}10^{9}$ & $7.2{\times}10^{9}$ & $9.3{\times}10^{9}$ & ${1.7{\times}10^{10}}^{\ast}$ & ${2.3{\times}10^{10}}^{\ast}$\\
  \addlinespace
  Diamond & $\lambda$ & $92.1$ & $104.5$ & $114.1$ & $125.6$ & $142.3^{\ast}$ & $157.3^{\ast}$\\
  & $\lambda$ (SG) & $95.2$ & $102.0$ & $106.8$ & $112.5$ & $120.1^{\ast}$ & $126.7^{\ast}$\\
  & Toffolis per walk & $5.9{\times}10^{5}$ & $8.8{\times}10^{5}$ & $1.3{\times}10^{6}$ & $1.6{\times}10^{6}$ & $2.5{\times}10^{6}$ & $3.5{\times}10^{6}$\\
  & Toffolis per walk (SG) & $1.6{\times}10^{5}$ & $2.8{\times}10^{5}$ & $3.2{\times}10^{5}$ & $4.0{\times}10^{5}$ & $6.7{\times}10^{5}$ & $8.6{\times}10^{5}$\\
  & Logical qubits & $4{,}284$ & $4{,}291$ & $8{,}393$ & $8{,}400$ & $16{,}598$ & $16{,}607$\\
  & Logical qubits (SG) & $4{,}284$ & $4{,}291$ & $8{,}393$ & $8{,}400$ & $16{,}598$ & $16{,}605$\\
  & Total Toffolis & $2.3{\times}10^{10}$ & $3.9{\times}10^{10}$ & $6.1{\times}10^{10}$ & $8.8{\times}10^{10}$ & ${1.5{\times}10^{11}}^{\ast}$ & ${2.3{\times}10^{11}}^{\ast}$\\
  & Total Toffolis (SG) & $6.6{\times}10^{9}$ & $1.2{\times}10^{10}$ & $1.5{\times}10^{10}$ & $1.9{\times}10^{10}$ & ${3.5{\times}10^{10}}^{\ast}$ & ${4.7{\times}10^{10}}^{\ast}$\\
  \addlinespace
  MgO & $\lambda$ & $143.1$ & $145.6$ & $146.3$ & $149.4$ & $151.6^{\ast}$ & $153.8^{\ast}$\\
  & Toffolis per walk & $6.0{\times}10^{5}$ & $8.9{\times}10^{5}$ & $1.3{\times}10^{6}$ & $1.6{\times}10^{6}$ & $2.6{\times}10^{6}$ & $3.5{\times}10^{6}$\\
  & Toffolis per walk (SG) & $1.6{\times}10^{5}$ & $2.7{\times}10^{5}$ & $3.1{\times}10^{5}$ & $3.9{\times}10^{5}$ & $6.7{\times}10^{5}$ & $8.7{\times}10^{5}$\\
  & Logical qubits & $4{,}284$ & $4{,}291$ & $8{,}393$ & $8{,}400$ & $16{,}598$ & $16{,}607$\\
  & Logical qubits (SG) & $4{,}284$ & $4{,}291$ & $8{,}393$ & $8{,}400$ & $16{,}598$ & $16{,}607$\\
  & Total Toffolis & $3.6{\times}10^{10}$ & $5.5{\times}10^{10}$ & $7.9{\times}10^{10}$ & $1.0{\times}10^{11}$ & ${1.7{\times}10^{11}}^{\ast}$ & ${2.3{\times}10^{11}}^{\ast}$\\
  & Total Toffolis (SG) & $9.7{\times}10^{9}$ & $1.7{\times}10^{10}$ & $2.0{\times}10^{10}$ & $2.5{\times}10^{10}$ & ${4.4{\times}10^{10}}^{\ast}$ & ${5.7{\times}10^{10}}^{\ast}$\\
  \addlinespace
  BN & $\lambda$ & $81.3$ & $92.2$ & $98.7$ & $106.7$ & $117.7^{\ast}$ & $127.6^{\ast}$\\
  & Toffolis per walk & $5.9{\times}10^{5}$ & $8.8{\times}10^{5}$ & $1.3{\times}10^{6}$ & $1.6{\times}10^{6}$ & $2.5{\times}10^{6}$ & $3.5{\times}10^{6}$\\
  & Toffolis per walk (SG) & $2.1{\times}10^{5}$ & $3.0{\times}10^{5}$ & $4.4{\times}10^{5}$ & $5.6{\times}10^{5}$ & $7.9{\times}10^{5}$ & $1.0{\times}10^{6}$\\
  & Logical qubits & $4{,}284$ & $4{,}291$ & $8{,}393$ & $8{,}400$ & $16{,}598$ & $16{,}605$\\
  & Logical qubits (SG) & $4{,}284$ & $4{,}291$ & $8{,}393$ & $8{,}400$ & $16{,}598$ & $16{,}605$\\
  & Total Toffolis & $2.1{\times}10^{10}$ & $3.5{\times}10^{10}$ & $5.3{\times}10^{10}$ & $7.4{\times}10^{10}$ & ${1.3{\times}10^{11}}^{\ast}$ & ${1.9{\times}10^{11}}^{\ast}$\\
  & Total Toffolis (SG) & $7.4{\times}10^{9}$ & $1.2{\times}10^{10}$ & $1.9{\times}10^{10}$ & $2.5{\times}10^{10}$ & ${4.0{\times}10^{10}}^{\ast}$ & ${5.7{\times}10^{10}}^{\ast}$\\
  \addlinespace
  AlN & $\lambda$ & $67.2$ & $71.2$ & $75.3$ & $78.2^{\ast}$ & $83.3^{\ast}$ & $87.5^{\ast}$\\
  & Toffolis per walk & $1.7{\times}10^{6}$ & $2.5{\times}10^{6}$ & $3.5{\times}10^{6}$ & $4.7{\times}10^{6}$ & $7.0{\times}10^{6}$ & $1.0{\times}10^{7}$\\
  & Toffolis per walk (SG) & $6.3{\times}10^{5}$ & $8.9{\times}10^{5}$ & $1.1{\times}10^{6}$ & $2.0{\times}10^{6}$ & $2.8{\times}10^{6}$ & $3.7{\times}10^{6}$\\
  & Logical qubits & $8{,}385$ & $8{,}392$ & $16{,}590$ & $16{,}599$ & $32{,}989$ & $32{,}996$\\
  & Logical qubits (SG) & $8{,}385$ & $8{,}392$ & $16{,}590$ & $16{,}599$ & $32{,}989$ & $32{,}996$\\
  & Total Toffolis & $4.7{\times}10^{10}$ & $7.7{\times}10^{10}$ & $1.1{\times}10^{11}$ & ${1.6{\times}10^{11}}^{\ast}$ & ${2.5{\times}10^{11}}^{\ast}$ & ${3.8{\times}10^{11}}^{\ast}$\\
  & Total Toffolis (SG) & $1.8{\times}10^{10}$ & $2.7{\times}10^{10}$ & $3.6{\times}10^{10}$ & ${6.6{\times}10^{10}}^{\ast}$ & ${9.9{\times}10^{10}}^{\ast}$ & ${1.4{\times}10^{11}}^{\ast}$\\
  \addlinespace
  CdSe & $\lambda$ & $69.3$ & $75.1$ & $78.5$ & $82.4^{\ast}$ & $88.5^{\ast}$ & $93.5^{\ast}$\\
  & Toffolis per walk & $9.6{\times}10^{5}$ & $1.5{\times}10^{6}$ & $2.0{\times}10^{6}$ & $2.7{\times}10^{6}$ & $4.1{\times}10^{6}$ & $5.7{\times}10^{6}$\\
  & Toffolis per walk (SG) & $3.2{\times}10^{5}$ & $4.4{\times}10^{5}$ & $7.4{\times}10^{5}$ & $9.3{\times}10^{5}$ & $1.3{\times}10^{6}$ & $1.7{\times}10^{6}$\\
  & Logical qubits & $4{,}284$ & $8{,}386$ & $16{,}586$ & $16{,}593$ & $32{,}983$ & $32{,}990$\\
  & Logical qubits (SG) & $8{,}379$ & $8{,}386$ & $16{,}586$ & $16{,}593$ & $32{,}983$ & $32{,}990$\\
  & Total Toffolis & $2.8{\times}10^{10}$ & $4.7{\times}10^{10}$ & $6.8{\times}10^{10}$ & ${9.4{\times}10^{10}}^{\ast}$ & ${1.6{\times}10^{11}}^{\ast}$ & ${2.3{\times}10^{11}}^{\ast}$\\
  & Total Toffolis (SG) & $9.4{\times}10^{9}$ & $1.4{\times}10^{10}$ & $2.5{\times}10^{10}$ & ${3.3{\times}10^{10}}^{\ast}$ & ${5.0{\times}10^{10}}^{\ast}$ & ${7.0{\times}10^{10}}^{\ast}$\\
\end{tabular}
\end{ruledtabular}
\caption{\textbf{BSE resources across $k$-point meshes.}
Triexcitonic ($m=3$) estimates unless labelled, with space-group adaptation only in the (SG) rows; the $10\times10\times10$ column is Fig.~\ref{fig:optimized-exciton-comparison}.
Diamond (SG) totals use the space-group-adapted fitted subnormalizations; the other materials retain the subnormalizations without space-group adaptation.
Subnormalizations $\lambda$ are in $E_h$.
Asterisks mark subnormalizations, and the resulting Toffoli counts, obtained from a power-law fit to the total subnormalization at the last three evaluated meshes, as in Fig.~\ref{fig:optimized-channel-scaling}(a).
Walk costs are evaluated at every mesh, and logical qubits are total counts: the $2m$ particle registers (momentum, band, and spin), the LCU control, the THC-index extension of the two routed registers, two exchange qubits (spin-projector signal and state-preparation flag), the peak primitive workspace, and the phase-estimation register of $2\lceil\log_2(\mathcal I+1)\rceil-1$ qubits for $\mathcal I$ walk steps~\cite{lee2021even}.}
\label{tab:material-resources}
\end{table*}

\begin{table*}[t]
\centering\small
\setlength{\tabcolsep}{4pt}
\begin{ruledtabular}
\begin{tabular}{llrrrrrr}
  \textbf{System} & & $4\times4$ & $5\times5$ & $6\times6$ & $7\times7$ & $16\times16$ & $32\times32$\\
  \hline
  MoSe$_2$ & $\lambda$ & $128.5$ & $134.6$ & $138.6$ & $139.9$ & $154.7^{\ast}$ & $167.8^{\ast}$\\
  & Toffolis per walk & $1.0{\times}10^{6}$ & $1.3{\times}10^{6}$ & $1.5{\times}10^{6}$ & $1.8{\times}10^{6}$ & $4.1{\times}10^{6}$ & $8.3{\times}10^{6}$\\
  & Toffolis per walk (SG) & $4.0{\times}10^{5}$ & $4.8{\times}10^{5}$ & $5.8{\times}10^{5}$ & $6.4{\times}10^{5}$ & $1.3{\times}10^{6}$ & $2.5{\times}10^{6}$\\
  & Logical qubits & $4{,}283$ & $8{,}385$ & $8{,}392$ & $8{,}392$ & $16{,}599$ & $32{,}996$\\
  & Logical qubits (SG) & $4{,}283$ & $8{,}385$ & $8{,}392$ & $8{,}392$ & $16{,}599$ & $32{,}996$\\
  & Total Toffolis & $5.7{\times}10^{10}$ & $7.5{\times}10^{10}$ & $9.1{\times}10^{10}$ & $1.1{\times}10^{11}$ & ${2.7{\times}10^{11}}^{\ast}$ & ${5.9{\times}10^{11}}^{\ast}$\\
  & Total Toffolis (SG) & $2.2{\times}10^{10}$ & $2.8{\times}10^{10}$ & $3.4{\times}10^{10}$ & $3.8{\times}10^{10}$ & ${8.8{\times}10^{10}}^{\ast}$ & ${1.8{\times}10^{11}}^{\ast}$\\
\end{tabular}
\end{ruledtabular}
\caption{\textbf{BSE resources for monolayer MoSe$_2$ across $k\times k$ meshes.}
Triexcitonic ($m=3$) estimates with the settings of Table~\ref{tab:material-resources}; the $32\times32$ column is Fig.~\ref{fig:optimized-exciton-comparison}.
The direct term uses the bare Coulomb interaction.
Subnormalizations $\lambda$ are in $E_h$.
Asterisks mark subnormalizations, and the resulting Toffoli counts, obtained from a power-law fit to the total subnormalization at the last three evaluated meshes ($5\times5$ to $7\times7$).
The (SG) rows combine the space-group-adapted walk with the subnormalization from the fit without space-group adaptation.
Walk costs and logical qubits are defined as in Table~\ref{tab:material-resources}.}
\label{tab:material-resources-2d}
\end{table*}

\end{document}